\documentclass[a4paper]{aa}
\usepackage{amsmath}
\usepackage{txfonts}
\usepackage{graphicx}
\usepackage{natbib}
\usepackage{multirow}
\usepackage{color}
\usepackage{array}
\usepackage{ifthen, hyperref}
\usepackage{fixltx2e}
\usepackage{caption}
\usepackage{lscape}
\usepackage{tabularx}
\usepackage{breakurl}
\usepackage{enumitem}
\bibpunct{(}{)}{;}{a}{}{,}

\begin{document}

\title{A novel method for transient detection in high-cadence optical surveys}

\subtitle{Its application for a systematic search for novae in M31}
\author{Monika~D.~Soraisam\thanks{{Present address: NOAO, Tucson, AZ, USA} \newline\email{monikas@mpa-garching.mpg.de}} \inst{\ref{MPA}}
    \and Marat~Gilfanov \inst{\ref{MPA},\ref{SRI}}
    \and Thomas~Kupfer \inst{\ref{Caltech}}
    \and Frank~Masci \inst{\ref{IPAC}}
    \and Allen~W.~Shafter \inst{\ref{SDU}}
    \and Thomas~A.~Prince \inst{\ref{Caltech}}
    \and Shrinivas~R.~Kulkarni \inst{\ref{Caltech}}
    \and Eran~O.~Ofek \inst{\ref{WIS}}
    \and Eric~Bellm \inst{\ref{Caltech}}
    }
\institute{Max Planck Institute for Astrophysics, Karl-Schwarzschild-Str.~1, 85748 Garching, Germany\label{MPA}
  \and Space Research Institute, Russian Academy of Sciences, Profsoyuznaya~84/32, 117997 Moscow, Russia\label{SRI}
  \and Division of Physics, Mathematics, and Astronomy, California Institute of Technology, Pasadena, CA 91125, USA\label{Caltech}
  \and Infrared Processing and Analysis Center, California Institute of Technology, Pasadena, CA 91125, USA\label{IPAC}
  \and Department of Astronomy, San Diego State University, San Diego, CA 92182, USA\label{SDU}
  \and Benoziyo Center for Astrophysics, Weizmann Institute of Science, 76100 Rehovot, Israel\label{WIS}
    }

\date{Received date / Accepted date}

\abstract
{In the present era of large-scale surveys in the time domain, the processing of the data, from procurement up to the detection of sources, is generally automated. One of the main challenges in the  astrophysical analysis of their output is contamination by artifacts, especially in the regions of high surface brightness of unresolved emission.
}
{We present a novel method for identifying candidates for variable and transient sources from the outputs of optical time-domain surveys' data pipelines. We use the method to conduct a systematic search for novae in the intermediate Palomar Transient Factory (iPTF) observations of the bulge part of M31 during the second half of 2013.}
{
We demonstrate that a significant fraction of artifacts produced by the iPTF pipeline form a locally uniform background of false detections approximately obeying Poissonian statistics, whereas genuine variable and transient sources as well as artifacts associated with bright stars result in clusters of  detections, whose spread is determined by the source localization accuracy. This makes the problem analogous to source detection on images produced by grazing incidence X-ray telescopes, enabling one to utilize the arsenal of powerful tools developed in X-ray astronomy. In particular, we use a wavelet-based source detection algorithm from the {\it Chandra} data analysis package \texttt{CIAO}.
}
{
Starting from $\sim 2.5\cdot 10^{5}$ raw detections made by the iPTF data pipeline, we obtain $\approx 4000$ unique source candidates.  Cross-matching these candidates with the source-catalog of a deep reference image of the same field, we find counterparts for $\sim 90\%$ of the candidates. These sources are either artifacts due to imperfect PSF matching or genuine variable sources. The remaining $\sim 400$ detections are transient sources. 
We identify novae among these candidates by applying selection cuts to their lightcurves based on the expected properties of novae.
Thus, we recovered all 12 known novae (not counting one that erupted toward the end of the survey) registered during the time span of the survey and discovered three nova candidates. Our method is generic and can be applied for mining any target out of the artifacts in optical time-domain data. As it  is fully automated, its incompleteness can be accurately computed and corrected for.}
{}
{}

\titlerunning{A novel method for optical transient detection}
\authorrunning{Soraisam et al.}
\maketitle

\section{Introduction}
\label{intro_5}

\begin{figure*}
\centering
\includegraphics[width=150mm]{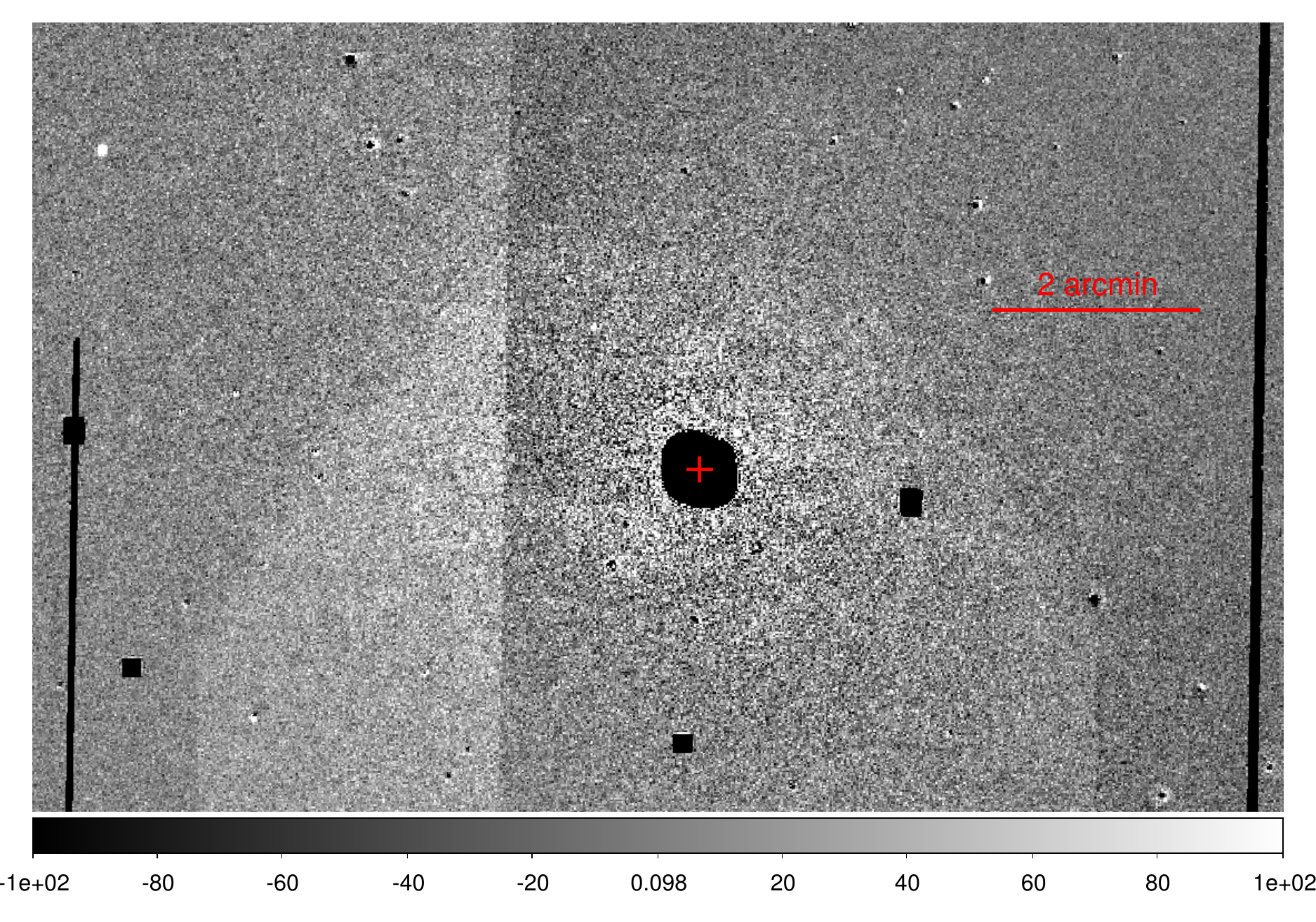}
\vspace{5mm}
\includegraphics[width=150mm]{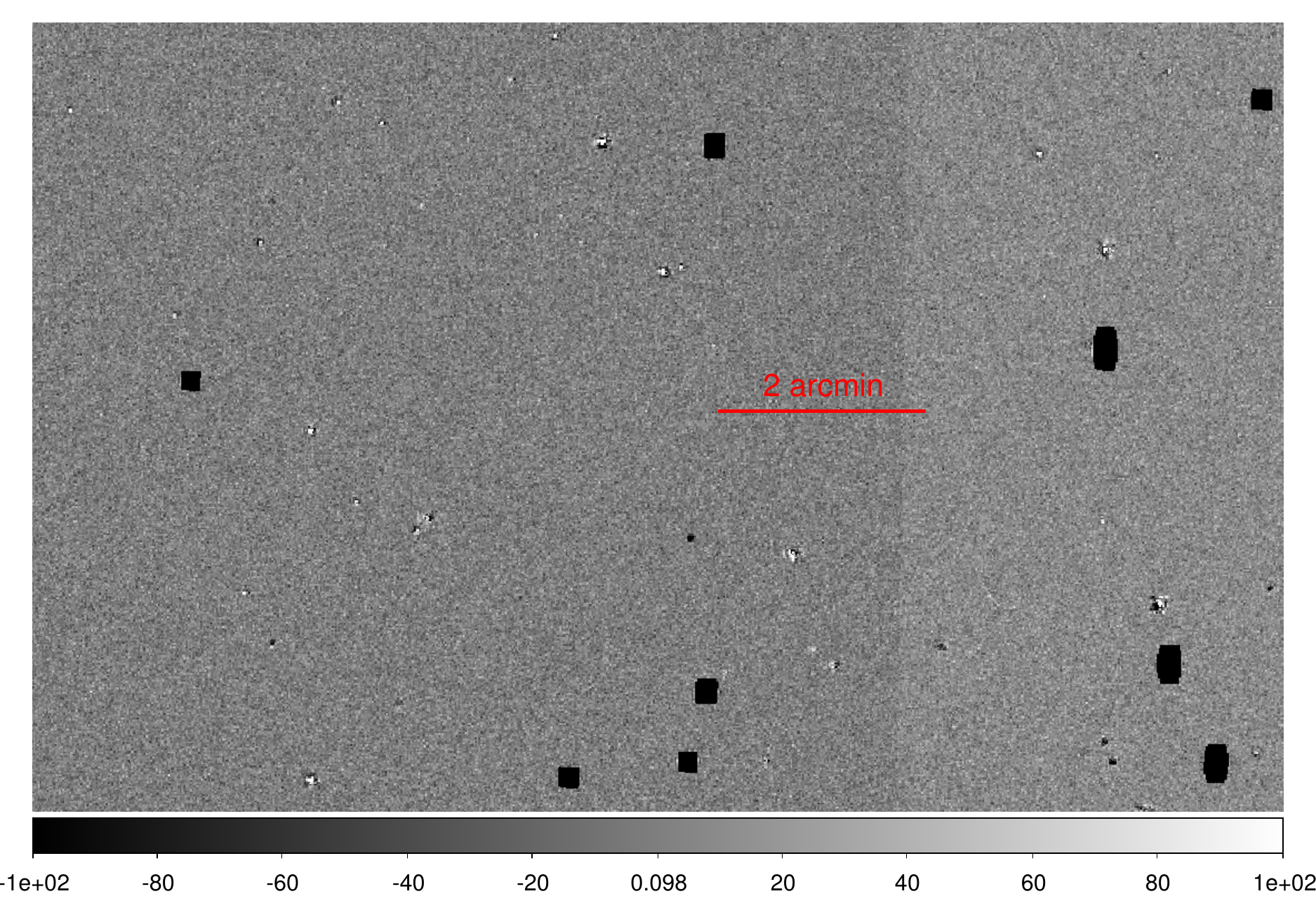} 
\caption{{\it Upper panel:} A section of a typical M31 difference image from the iPTF pipeline containing the bulge. {\it Lower panel:} The outer part of the same difference image, about half a degree northward of the center of M31. The masked areas are assigned large negative values. North is upward and east is left. The center of M31 is marked by the red cross in the upper panel. The vertical sharp edges seen in both panels arise from the boundary of blocks that have been used to match the background before the image subtraction in the iPTF DI pipeline.}
\label{bad-img}
\end{figure*}

In time-domain astronomy, the difference imaging (DI) technique \citep{Tomaney-1996, Alard-1998, Alard-2000, Alcock-1999, Wozniak-2000, Bond-2001, Gal-Yam-2008} provides an efficient way to detect flux transients. 
In particular, as compared to the former conventional method of catalog cross-matching, DI provides a more effective strategy for finding variable and transient sources in crowded stellar fields. Such fields, as difficult as they are to analyze, hold equally rich opportunities for astrophysical studies. 
DI has thus become the staple choice as is evidenced by its implementation in the majority of the data analysis pipelines of modern time-domain surveys, e.g., the intermediate Palomar Transient Factory (iPTF) survey (see \citealt{Masci-2016}).

The basic principle of DI is to match the point-spread function (PSF) and the background, both of which often vary spatially, between a reference image and an input/science image.  In the commonly used approach \citep[e.g.,][]{Alard-1998}, this match is accomplished by convolving one of the images, generally the reference image with better seeing and higher signal-to-noise ratio, with a kernel that minimizes the difference between the convolved reference image and the science image. The determination of this kernel forms the most important step in such form of DI technique. 
An alternative and statistically more rigorous approach to implementation of DI for transient detection was recently proposed by \citet{Zackay-2016}.
DI in its traditional formulation is implemented in many automated pipelines, particularly for large-scale surveys. 
Its execution, however, in almost all cases is subject to too many false positives/artifacts compromising the quality of the resulting difference image, especially for the automated case. 
There are several  reasons which  may make the difference images susceptible to artifacts --- the inherent instability of the mathematical process used (see \citealt{Zackay-2016} for details on this), registration errors, improper PSF and/or background matching, image edges, cosmic rays, etc. Consequently, the catalogs of sources detected in the difference images generally become contaminated, in fact dominated, by artifacts \citep[see for example][]{Bloom-2012}.

Furthermore, when dealing with a field containing a background that is spatially varying and bright, such as the bulge of M31, the DI quality tends to deteriorate further \citep{Kerins-2010}. To illustrate this, we zoom in on two parts of a typical difference image of M31 generated by the iPTF DI pipeline (\citealt{Masci-2016}), the bulge part and an outer part about half a degree northward of the center of M31, as shown in the upper and lower panels of Fig.~\ref{bad-img}, respectively. 
It is clearly evident comparing the two sections of the same difference image, that the residuals in the bulge part have much higher amplitudes than in the outer part. As such the bulge part is even more prone to artifacts. 
Thus, the artifact-contamination in the difference image catalogs for such fields worsens toward the bulge.

In this paper, using iPTF M31 observations as an example, we present a novel method to efficiently identify candidates for variable and transient sources in an artifact-dominated image and recover candidates even from the bulge, where the DI quality is low. We then make use of the method to conduct a systematic search for novae in M31. Although we describe our procedure using iPTF observations, the issues discussed above are generic and therefore our method is equally applicable to other surveys and other fields. 
The method involves mapping the sources detected from difference images onto a blank image of the corresponding field, thereby storing their spatial and recurrence information. We call the resulting image {\em spatial recurrence image}. In this image, the candidates for variable and transient sources appear as clusters of points over a background made up primarily of systematic artifacts, which are random and approximately follow Poisson distributions. In this latter aspect, the image becomes analogous to an X-ray image. Thus, exploiting the approximately Poissonian nature of the background, we make use of image analysis tools for source detection developed in X-ray astronomy, to obtain a list of unique candidates for variable and transient sources, significantly less contaminated by artifacts. In particular, we use the wavelet-based tool \texttt{WAVDETECT} \citep{Freeman-2002} from the {\em Chandra} Interactive Analysis of Observations (\texttt{CIAO}; \citealt{CIAO-2006}) software package. Once the candidates for variable and transient sources are obtained, we then follow the fairly standard practice of searching for novae --- we construct the lightcurves for the candidates and then apply cuts based on expected properties of novae.

The paper is organized as follows. In Sect.~\ref{iptf_data}, we describe the iPTF M31 data used in our analysis, and in Sect.~\ref{space_time}, we present the spatial recurrence image of the sources detected in the difference images of the M31 observations by the iPTF pipeline. In Sect.~\ref{method}, we describe the method we have developed to obtain the candidates for variable and transient sources from outputs of data pipelines of time-domain surveys. The procedure for searching novae amongst these candidates is outlined in Sect.~\ref{nova_search}. 
We show the results in Sect.~\ref{results} and end with their discussion and a summary in Sect.~\ref{discuss}.

\begin{figure*}
\centering
\includegraphics[width=150mm]{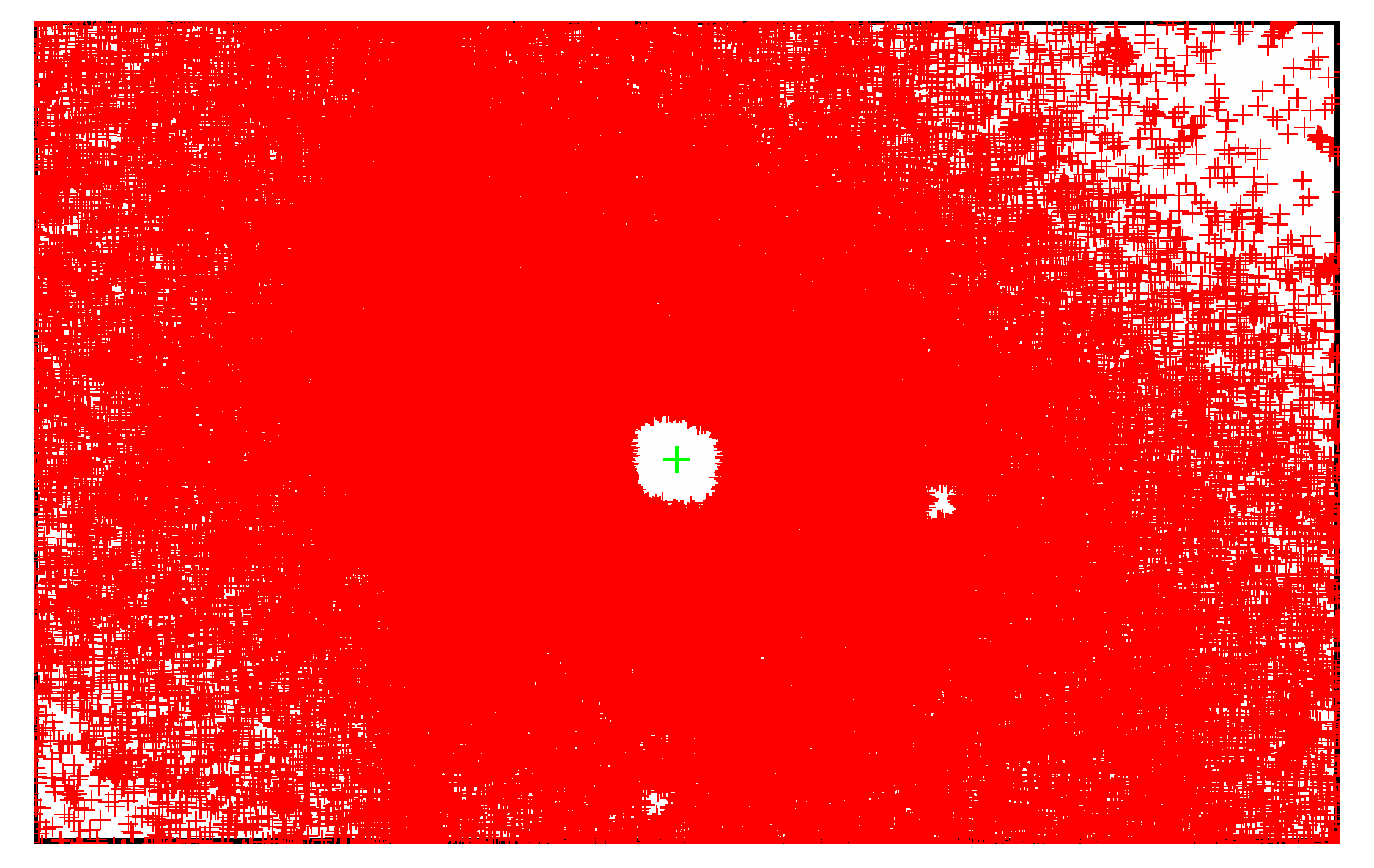}
\includegraphics[width=150mm]{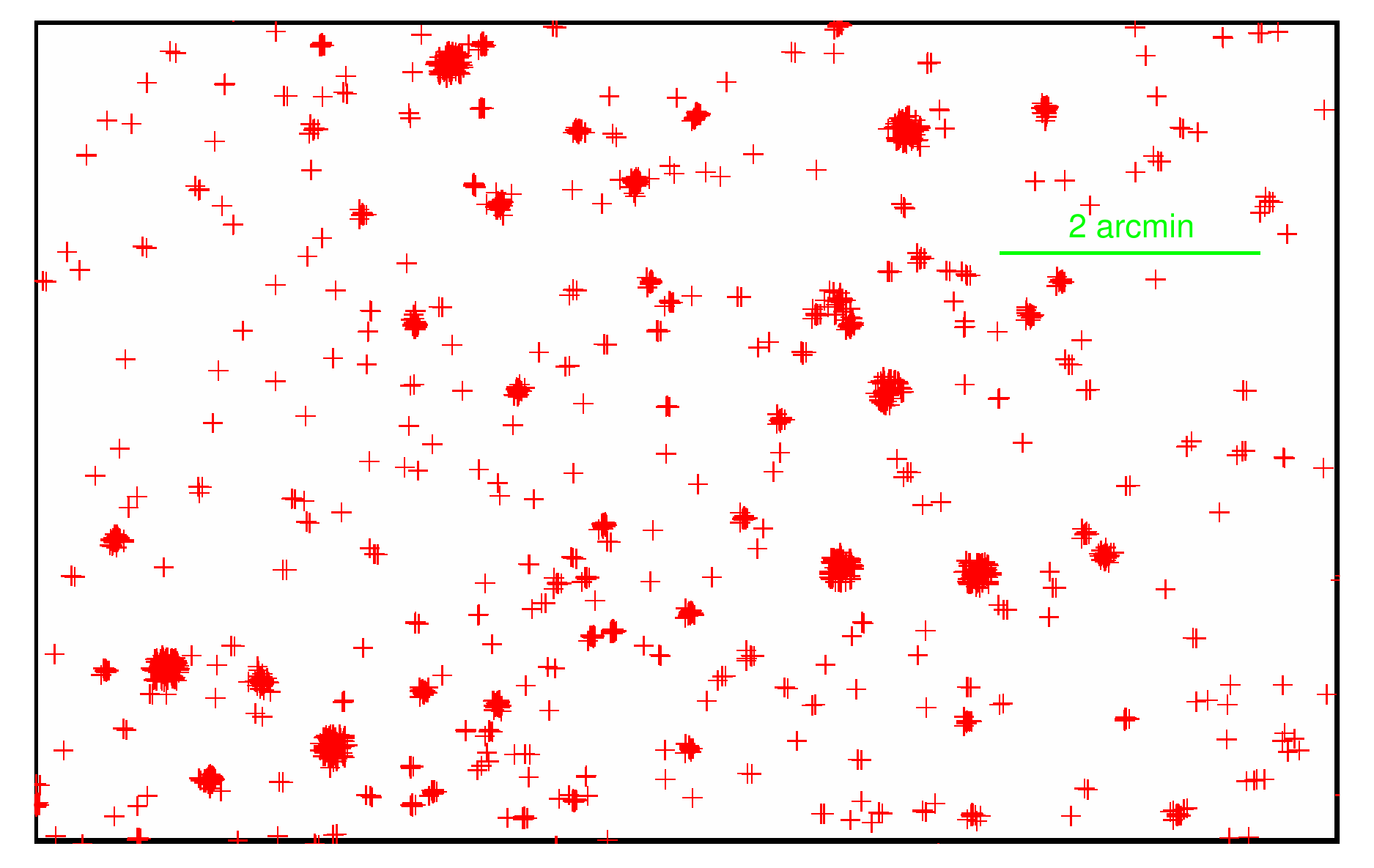}
\caption{{\it Upper panel:} The spatial recurrence image obtained by plotting the raw detections as red crosses (cf.~Sects.~\ref{intro_5}, \ref{space_time}), zoomed in on the bulge of M31. The center of M31 is marked by the green cross. North is upward and east is left. Due to saturation close to the center of M31, there are no sources detected by the iPTF pipeline. {\it Lower panel:} The outer part of the spatial recurrence image, about half a degree northward of the center of M31.}
\label{fig:cluster}
\end{figure*}

\begin{figure*}
\centering
\includegraphics[width=150mm]{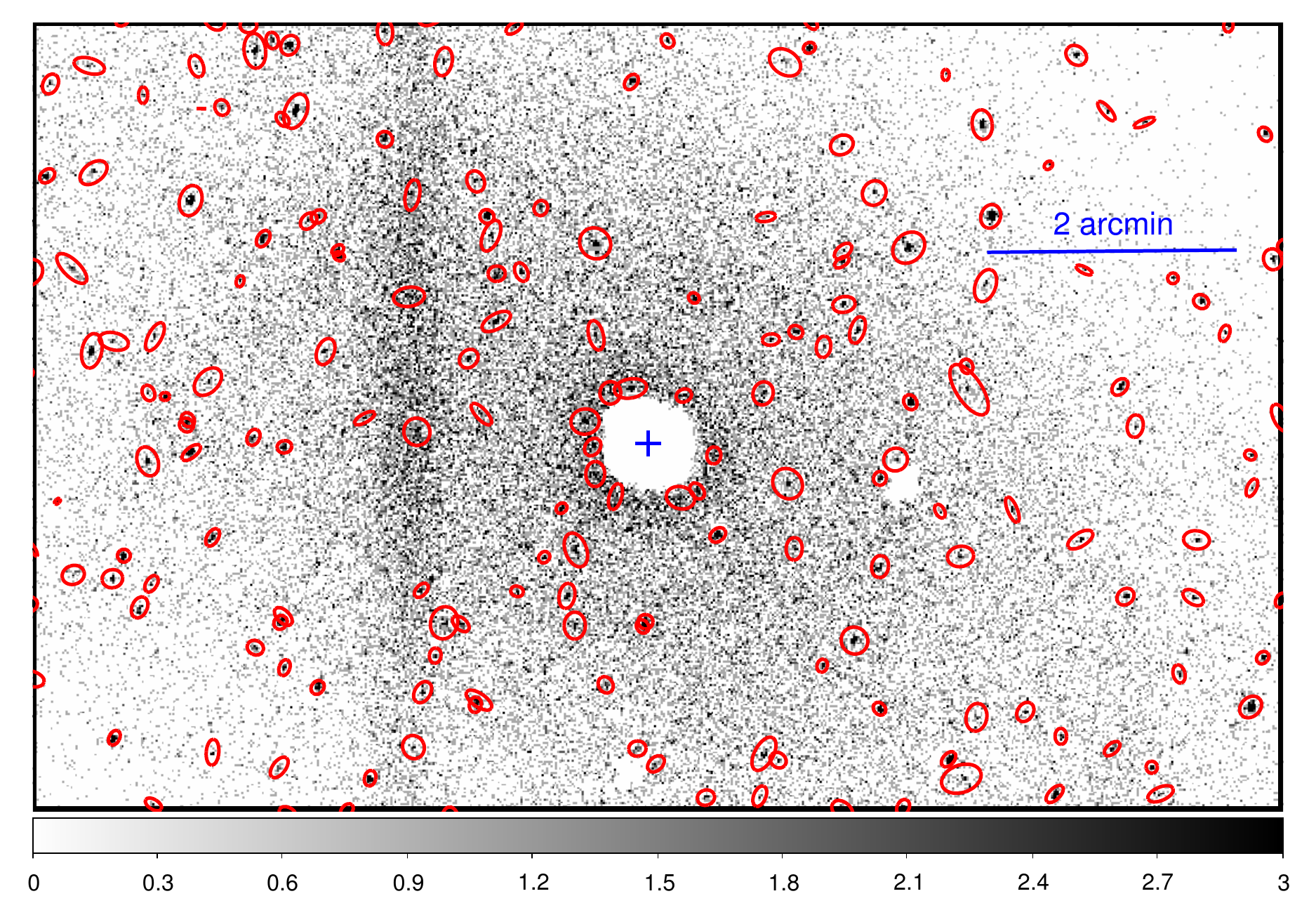}
\includegraphics[width=150mm]{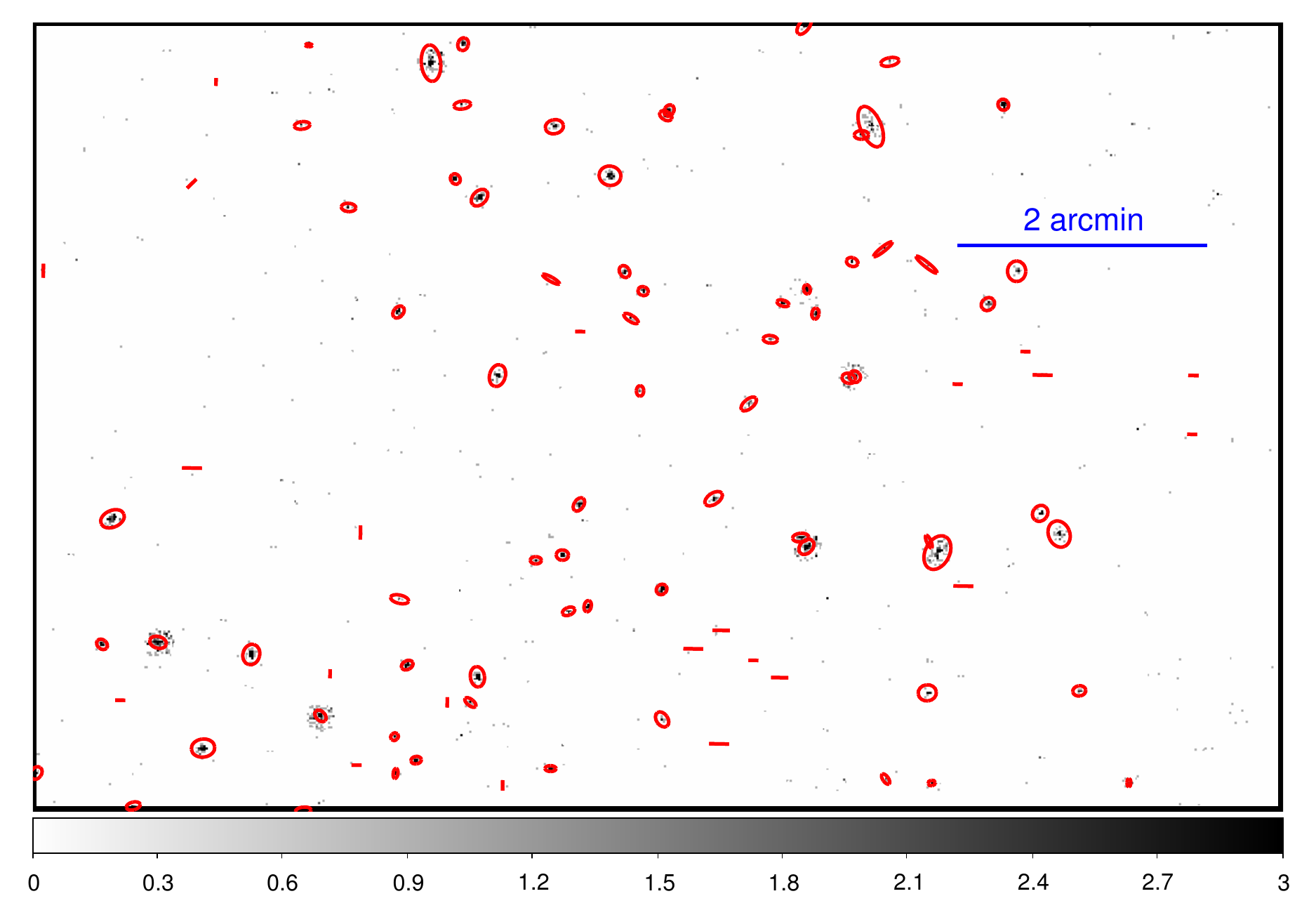}
\caption{{\it Upper panel:} The hits image obtained by counting the raw detections within each pixel (cf.~Sect.~\ref{space_time}), zoomed in on the bulge. The blue cross marks the center of M31 and the red ellipses show the location of sources found by \texttt{WAVDETECT} in this region. North is upward and east is left. As the area around the center is masked upstream in the iPTF pipeline due to saturation, the corresponding pixels in the hits image have zero values. The stripe that can be seen left of the center arises from the border between blocks used to match the background in the iPTF DI pipeline (cf.~Fig.~\ref{bad-img}). {\it Lower panel:} The outer part of the hits image, half a degree northward of the M31 center, with the red ellipses showing the location of sources found by \texttt{WAVDETECT}. The shape of the elliptical region characterizes the shape of the distribution of counts within the source cell. Some of these regions output by \texttt{WAVDETECT} may have a very small length along one of the axes, such that they appear like lines; these regions are however only meant for visualization (see text).}
\label{fig:hits}
\end{figure*}

\section{iPTF M31 data}\label{iptf_data}

The Palomar Transient Factory (PTF) survey \citep{Law, Rau, Ofek-2012} was succeeded by iPTF in 2013. Both share the same hardware and pipelines but some upgrades were implemented for the latter. The survey covers $\approx 7.26~{\rm deg}^{2}$ on the sky with the camera mounted on the 1.2~m Samuel Oschin Telescope at the Palomar Observatory. The observations are conducted primarily in the $R$ band, with some also made in the $g$ band. The pixel scale of the detector is $1.01''$ and the typical seeing is $\approx 2''$. With the nominal exposure time of 60~seconds, the $5\sigma$ limiting magnitude reaches $m_{R}\approx 21$.
The M31 field analyzed here, covering $\approx 1\times 0.5~{\rm deg}^{2}$ and including the bulge, was observed by iPTF between September 2013 and January 2014, comprising 201 epochs (all taken in the $R$ band).

DI was performed for these epochs by the iPTF DI pipeline, which among other things performs the image convolution for PSF-matching using a technique developed by the iPTF team (\citealt{Masci-2016}). The pipeline outputs as primary products photometric tables containing information (celestial coordinates, differential fluxes, etc.) on the sources detected (the detection criterion being signal-to-noise ratio $S/N\gtrsim 3.5$) in the difference images. Users are responsible for further characterization of bona fide sources and the subsequent selection of objects of interest. 
Given that these detections are not artifact-free and thus are to be further cleaned, we here refer to these detections as {\em raw detections}. The typical number of raw detections per epoch extracted from the difference images by the iPTF pipeline is $\sim 1000$, and the total for this season is $\sim 2.5\cdot10^{5}$. The latter value of course includes multiple detections from any single source.

\section{Spatial recurrence map of  raw detections}\label{space_time}
With the goal of identifying unique variable and transient sources, we construct a spatial recurrence map of all the $\sim 2.5\cdot10^{5}$ raw detections from Sect.~\ref{iptf_data}. To this end, we create a blank image of the same M31 field, sampling the pixels at the iPTF scale of $1.01''$ \citep{Law}. 
We then plot all the raw detections in this image, marking their positions with crosses.
The resulting image, zoomed in on the bulge and on the outer part, is shown in Fig.~\ref{fig:cluster} upper and lower panels, respectively.

The raw detections from individual sources cluster at the respective positions of the sources on the spatial recurrence image.
These clusters can be clearly seen in a sea of single raw detections in the outer part of the M31 field in Fig.~\ref{fig:cluster} lower panel. These give the locations of the candidates for variable sources, as well as transient sources appearing during the observation season, in the M31 field. This procedure thus provides a means for identifying unique variable and transient source candidates in one go, without the need for cross-matching the catalogs of raw detections from the DI pipeline. The bulge is, however, swamped by detections (see Fig.~\ref{fig:cluster} upper panel). This is the cumulative effect of the low-quality DI at the bulge and higher concentration of real sources (cf. Sect.~\ref{intro_5}). Particularly for large-scale surveys in the time domain, the bulge part of a galaxy is thus one of the main challenges, where the bulk of the candidates are lost using conventional methods of thresholding on the parameters of the raw detections or even via machine-learning algorithms. 
Our method presented in this paper improves on this limitation of recovering candidates from the central parts of galaxies.

To facilitate further analysis, we first convert the spatial recurrence image (Fig.~\ref{fig:cluster}) to a counts image, representing each raw detection as one count. 
In the resulting counts image, which we call here the {\em hits image}, the pixel value equals the total number of occurrences of a raw detection within the given pixel. This hits image is shown in Fig.~\ref{fig:hits}. Certainly some of the non-zero pixels in the image will be sources and the rest artifacts; the artifact contamination is expected to be more drastic in the bulge as can be inferred from the increased density of hits in that area (Fig.~\ref{fig:hits} upper panel).

\section{The method}
\label{method}
In this section, we describe our procedure for identifying candidates for variable and transient sources in the M31 field analyzed. Only minimal information, namely the positions of all raw detections in an epoch, is required for the implementation of our method, and this is conveniently stored in the form of the hits image (Fig.~\ref{fig:hits}). On this hits image, we perform a series of wavelet transformations using the \texttt{CIAO} tool \texttt{WAVDETECT}.

\subsection{Source detection using \texttt{WAVDETECT}}
\label{wavdetect}
\texttt{WAVDETECT} is one of the most popular tools for X-ray source detection from the \texttt{CIAO} software package, based on wavelet analysis. In brief, it correlates the Mexican-hat wavelet functions with the input image. At every pixel ({\it i,j}), the local background $B_{i,j}$ is estimated using the negative annulus of the wavelet. Based on a user-defined siginificance value $S$ (see Eq.~\ref{correlation}), \texttt{WAVDETECT} computes a threshold correlation value ($C_{i,j,0}$) for each pixel. It then identifies source pixels exhibiting correlation coefficients greater than their corresponding thresholds, and then these pixels are grouped into individual sources (see below) thus producing a final list of sources. Mathematically, \texttt{WAVDETECT} determines $C_{i,j,0}$ as 
\begin{equation}
 S=\int^{\infty}_{C_{i,j,0}}p(C|B_{i,j})dC.
 \label{correlation}
\end{equation}
Here $p(C|B_{i,j})$ is the probability of obtaining the correlation value $C$ for the given background $B_{i,j}$ at the pixel ($i,j$), in the absence of a source in this pixel. This probability distribution has been determined from simulations (see \citealt{Freeman-2002} for details), and the results are already encoded within the software.
If the correlation value $C_{i,j}$ at pixel ($i,j$) is greater than $C_{i,j,0}$, then it passes through as a source pixel. 

The correlation is performed at multiple wavelet scales defined by the user, where small (large) scales are sensitive to small (large) features. These scales $s$ define the size of the wavelet; the wavelet crosses zero at $\sqrt{2}s$ and extends effectively up to $5s$. 
The sources manifest themselves as islands (also called source cells) with some net counts above a background of nearly zero values when the image is smoothed with the positive part of the wavelet at the different scales and the background subtracted. The detected pixels are then assigned to their corresponding sources, and the final list of sources is made.
The two parameters, namely the significane value/threshold and the wavelet scales, are thus the main inputs for source detection in \texttt{WAVDETECT}.

In processing the hits image via \texttt{WAVDETECT}, we adopt nine different scales ranging from 1 to 16 pixels, in multiplicative steps of $\sqrt{2}$. For the threshold, we choose a value of $S=10^{-6}$. The M31 field analyzed here encompasses $\sim 10^{6}$ pixels and therefore the above threshold implies an expectation value of $\sim$ one false positive if the model for $p(C|B_{i,j})$ is accurate \citep[see][]{Freeman-2002}.

Our \texttt{WAVDETECT} run found 3981 sources in the hits image. The upper and lower panels of Fig.~\ref{fig:hits} show the sources found in the bulge and half a degree northward of the center of M31, respectively, which are marked by the red ellipses. It is to be noted that the elliptical regions generated by \texttt{WAVDETECT} are only meant for visualization of the source locations, and do not affect the subsequent operation of \texttt{WAVDETECT} (\texttt{CIAO}; \url{http://cxc.harvard.edu/ciao/download/doc/detect_manual/wav_ref.html}).

The interpretation of the \texttt{WAVDETECT} result is that there are 3981 {\em different} locations from which multiple detections of the signal on the difference images occurred. Obviously, these locations are candidates for variable and transient sources in the M31 field obtained from the five-month-long iPTF season. This number may still be contaminated by artifacts due to statistical fluctutaions that are not modeled accurately by the model for $p(C|B_{i,j})$. However, the number is largely free of the background of single raw detections (cf.~Sect.~\ref{space_time}, Fig.~\ref{fig:cluster} lower panel), as shown in Sect.~\ref{wave_counts}. Of course, some of these single raw detections may be genuine sources. 
 However, characterization of the nature of these single detections would require individual inspection of each of them, which is not feasible given the large number of artifacts produced by the iPTF data pipeline (but see \citealt{Zackay-2016} for an alternative approach to DI implementation).

\begin{figure*}
\centering
\includegraphics[width=88mm]{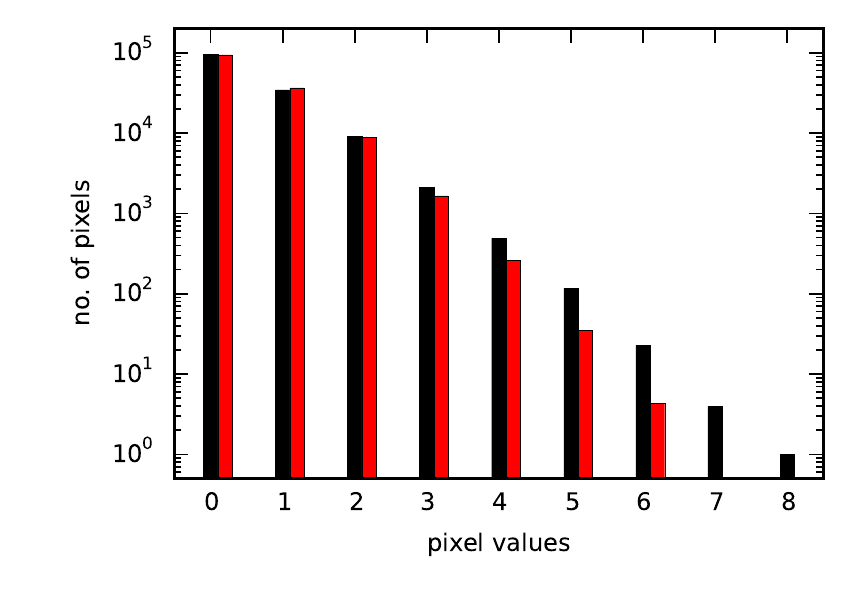} \hfill \includegraphics[width=88mm]{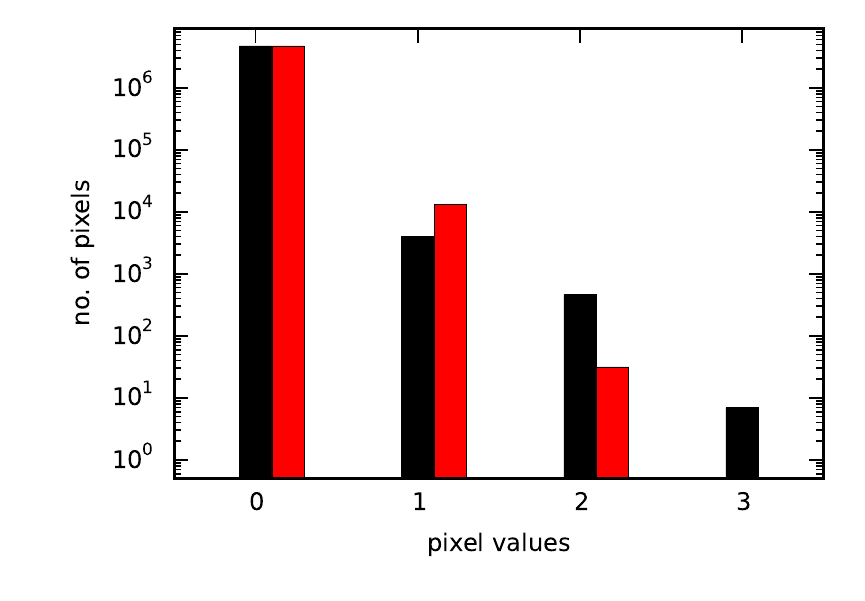}
\caption{Comparison of the background pixel value distribution of the hits image, shown in black, with the distribution expected for Poissonian background, shown in red, obtained using the \texttt{WAVDETECT} output $B'$ (see Sect.~\ref{sec:poisson}), for the bulge ({\it left}) and for the disc ({\it right}). The red histograms have been shifted along the x-axis by pixel value=0.1 for clarity.}
\label{fig:poisson}
\end{figure*}

\subsection{Applicability of \texttt{WAVDETECT} to the hits image}
\label{sec:poisson}

In the formulation of the \texttt{WAVDETECT} algorithm, the statistics of the background are tied to the probability distributions of the wavelet correlation values, $p(C|B_{i,j})$ in Eq.~\eqref{correlation}. 
The algorithm by itself is quite versatile as it can be adapted for any given statistical characteristic of the background by computing the corresponding distribution of $p(C|B_{i,j})$. However, in the implementation of \texttt{WAVDETECT} within the \texttt{CIAO} software package, Poisson statistics for the background are assumed, as it is designed for X-ray data. In general for the algorithm, the statistics of the sources themselves do not come in the picture; what matters is only that they stand out above the Poissonian fluctuations of the background.

The second assumption of \texttt{WAVDETECT} is that 
the expectation value of the background is constant \citep{Freeman-2002}. However, in real X-ray images, a constant background is definitely an idealization.
In the execution of \texttt{WAVDETECT}, the background at a given pixel of the image is estimated from the region covered by the negative annulus of the wavelet centered at the pixel. 
Thus, the assumption of a constant expectation value for the background is effectively only made for a region of radius $5s$ ($s$ is the wavelet scale) around each pixel.
Therefore, as long as there are no sharp variations in the background, the efficacy of \texttt{WAVEDETECT} is not affected \citep{Freeman-2002}.

When using \texttt{WAVDETECT} on the hits image, we are thus implicitly assuming the above conditions about its background --- that it follows Poisson statistics and that the mean values do not vary sharply. In order to analyze how accurate these assumptions are, we use the background image (let us call it $B'$) computed by \texttt{WAVDETECT}. In particular, we compare the pixel value distribution of the actual background of the hits image with that expected based on $B'$.
To this end, we mask out circular areas, each of radius $10''$ ($\approx 10$ pixels), centered on the sources detected by \texttt{WAVDETECT}.  
For each of the remaining background pixels, we compute the expected distribution of counts in that pixel assuming Poisson distribution with the mean given by $B'_{i,j}$.
We then sum these distributions and normalize the result to the total number of pixels in the sources-masked hits image. The resulting distribution is then compared with the actual pixel value distribution of the hits image with masked sources. As discussed in Sects.~\ref{intro_5} and \ref{space_time}, the DI quality is systematically inferior in the bulge as compared to the disc (Fig.~\ref{bad-img}). 
We thus make the above comparison of the pixel value distributions separately for the bulge and disc. We define the bulge of the M31 field as an elliptical region of semi-major axis $6'$ and axis ratio 0.47, centered on the M31 nucleus. The results are shown in the left and right panels of Fig.~\ref{fig:poisson}, respectively.

As can be seen from Fig~\ref{fig:poisson}, the background of the hits image does not strictly comply with Poissonian statistics, but the deviations are not stark. Moreover, even if the background were strictly Poissonian, one might expect that there could be some sources that barely stand out above the background. Those sources may not be detected above the threshold of $S=10^{-6}$ that we have applied. Thus, these sources would not be masked and they would distort the observed histograms in Fig.~\ref{fig:poisson}. So even if the true background is exactly Poissonian and with a constant expectation value, the two histograms may not look precisely the same. 
In conclusion, the application of \texttt{WAVDETECT} on the hits image is justified to the accuracy sufficient for the purpose of this analysis.

\subsection{Characteristics of \texttt{WAVDETECT} detections in the time-domain context}
\label{wave_counts}
It is quite evident from the range in the distribution of the background pixel values of the hits image (Fig.~\ref{fig:poisson}) that the detection threshold ($C_{i,j,0}$ in Eq.~\ref{correlation}) must be higher in the bulge than in the disc.

\begin{figure}
\centering
\includegraphics[width=88mm]{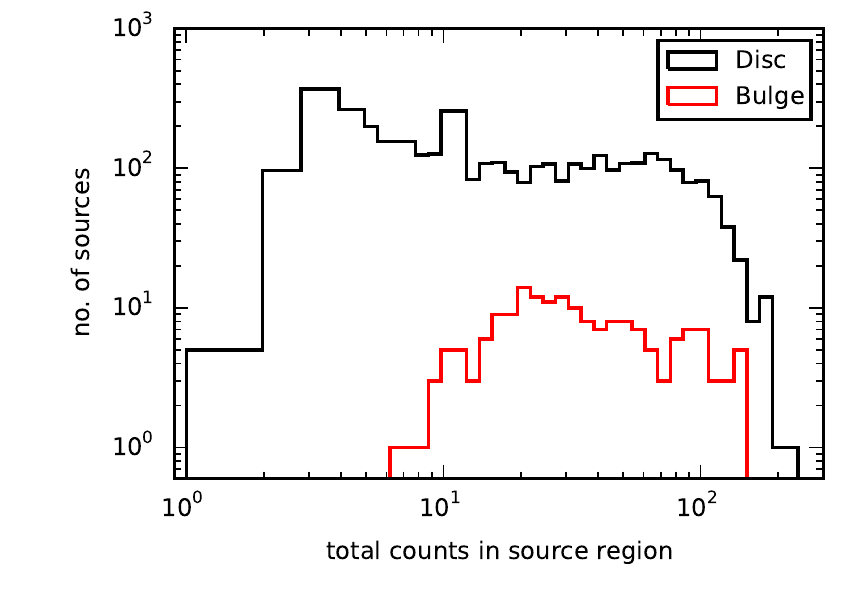}
\caption{Distribution of counts (i.e., the number of hits) in the source regions for the detections made by \texttt{WAVDETECT} in the bulge and the disc of the iPTF M31 field analyzed here. The five sources in the disc with one count passed through \texttt{WAVDETECT} due to their location in extensive empty areas.}
\label{fig:threshold_counts}
\end{figure}

Besides the positions of the detected sources, \texttt{WAVDETECT} also includes in its output their {\em counts} --- the sum of the pixel values in the corresponding source cells (see Sect.~\ref{wavdetect}). In the context of the hits image, the source counts have the meaning of the number of times a signal has been detected on the difference images from the given  position.  It is known that source counts computed by \texttt{WAVDETECT} are only crude estimates, since the tool is designed largely for detection and not for accurate photometry (\texttt{CIAO}). Of course, the former is our sole objective for use of the \texttt{WAVDETECT} tool. 
These crude source counts nevertheless allow us to gauge the difference in the sensitivity of \texttt{WAVDETECT} between the bulge and the disc of the M31 field, or in general between fields with sparse and crowded distributions of raw detections from the DI pipeline.
The distributions of source counts in the bulge and disc of the hits image are shown in Fig.~\ref{fig:threshold_counts}. 
As can be seen, the threshold number of hits for \texttt{WAVDETECT} detection is 7 in the bulge. In the disc, there are five sources detected with just one count; this arises due to these sources being located in extensive void areas, and this is a known feature of \texttt{WAVDETECT} (\texttt{CIAO}). However, they represent only $0.1\%$ of the total number of detected sources (5 out of 3981), and are thus insignificant. Effectively, the threshold count in the disc is 2.
The above discussion implies that, in the low background regions, \texttt{WAVDETECT} interprets source cells with two hits as sources; in the bulge region (Figs.~\ref{fig:cluster} upper panel and \ref{fig:hits}), due to the high density of artifacts, at least seven hits within a source cell are required for reliable detection. 
Note that these values are for the adopted significance of $S=10^{-6}$ in running \texttt{WAVDETECT} (Sect.~\ref{wavdetect}).

\begin{figure}
\centering
\includegraphics[width=88mm]{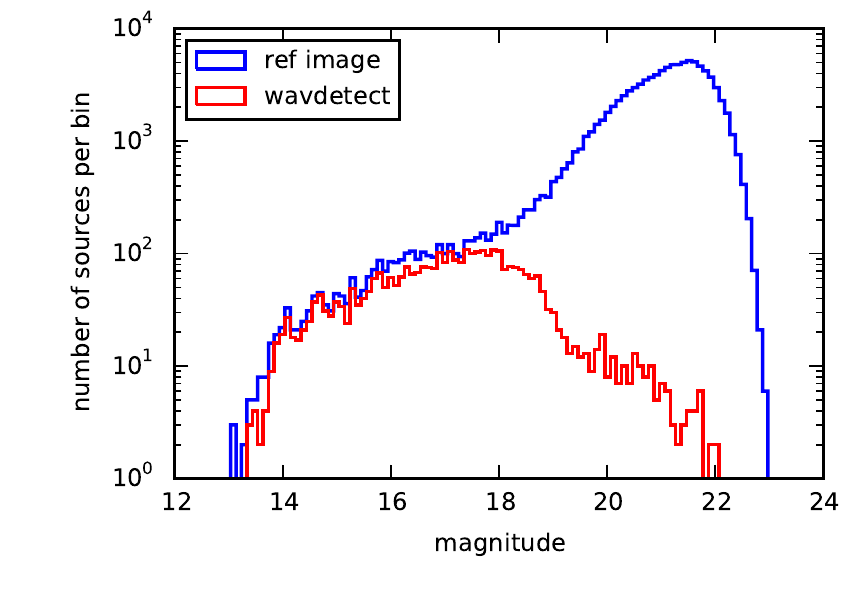}
\caption{Distribution of the magnitudes for the sources detected in the reference image and those \texttt{WAVDETECT} sources with counterparts in the reference image. The magnitudes of these \texttt{WAVDETECT} sources are measured from the reference image.}
\label{fig:reference_mag}
\end{figure}

As discussed in Sect.~\ref{wavdetect}, the 3981 sources detected by \texttt{WAVDETECT} in the hits image may not all be genuine sources. For instance, false positives may arise due to imperfect PSF-matching occurring persistently at the same position, for example at the position of a bright star. There may also be \texttt{WAVDETECT} false detections in the regions of high artifact density, for example caused by the deviations of the background distribution from being Poissonian. The latter is particularly relevant for those \texttt{WAVDETECT} sources at the low-count end of the distribution in Fig.~\ref{fig:threshold_counts}. 
To assess broadly the nature of possible artifact-contamination, in particular related to intrinsic factors such as imperfect PSF and/or background matching, we cross-match the \texttt{WAVDETECT} sources with the sources detected in a deep reference image (co-add of 24 images) of the same field of M31. 
Out of 3981, we obtain matches for 3525 \texttt{WAVDETECT} sources. A matching radius of $r\approx 2''$ --- the typical full width at half maximum (FWHM) of the iPTF point spread function --- was used in the cross-matching. 
The magnitude (in the $R$ band) distribution of the matched sources is shown in Fig.~\ref{fig:reference_mag}, along with that of all the sources in the reference image. Note that the magnitudes of the matched \texttt{WAVDETECT} sources are those measured in the reference image itself. Quite interestingly, the vast majority  of bright sources in the reference image are detected in the difference images of the iPTF pipeline. Among these matches, some fraction may be genuinely variable stars, but some may be due to systematics in the iPTF DI pipeline caused by, for example, imperfect PSF and background matching procedure.
Classification of artifacts from these matched sources is however beyond the scope of this work.

\begin{table*}
\captionsetup{justification=centering}
\caption{Summary of \texttt{WAVDETECT} detections}
\label{table:wave}
\renewcommand\arraystretch{1.5}
\centering
\begin{tabularx}{350pt}{X r}
\hline
Raw detections from iPTF pipeline (not neccessarily unique sources)	&254765\\
Number of unique sources obtained from \texttt{WAVDETECT}	&3981\\
Total number of raw detections encompassed by the \texttt{WAVDETECT} sources	&120287\\
Number of \texttt{WAVDETECT} sources with counterparts in the reference image	&3525\\
Expected number of false matches in the above	&60\\
Number of transient candidates	&456\\
\hline
\end{tabularx} 
\end{table*}

The total number of sources detected in the reference image down to $m_{R}\approx 23$ is 99041. Thus,  cross-matching the $\sim 4000$ \texttt{WAVDETECT} sources with the reference image source catalog may lead to some number of false matches. To compute the expected number of false matches $\left<N_{\rm f}\right>$, we use Eq.~\eqref{eq:false_match2} in Appendix~\ref{rand_match}.  
Assuming a uniform density of reference image sources in the bulge and disc separately, we compute $\left<N_{\rm f}\right>$ in the respective region. We then sum them and obtain the total number of expected false matches as $\approx 60$, which is about 2\% of the matched sources.

The remaining 456 \texttt{WAVDETECT} sources are then candidates for transients, i.e., objects without counterparts in the deep reference image (different from variable sources in this context). The classification of these sources is an ongoing project and is beyond the scope of this paper. The factors that can cause artifact-contamination for this subset, if any, may include image edges and imperfect background matching. The latter, however, has to occur persistently in the same location on the sky. This may happen due to imperfect filtering of detector and optical artifacts when creating the reference image used for the subtraction, thus contaminating all the resulting difference images obtained with the given reference image.

In Table~\ref{table:wave}, we summarize the characteristics of the \texttt{WAVDETECT} detections. For transients, our method thus scales down the search sample by nearly three orders of magnitude from the raw detections of $\sim 2.5\cdot 10^{5}$ to a sample of a few $10^{2}$ candidates. However, it is to be noted that the 60 false matches from above constitute about 12\% of the total number of transient candidates. Depending on the task at hand, one may choose to minimize the  number of candidates and consider only the 456 sources for characterization and follow-up studies or take a more conservative approach and take into consideration  all 3981 candidates. For example, the  number of candidates may be critical for the feasibility of follow-up spectroscopy, which is resource expensive. On the other hand, lightcurve filtering can be easily done for all 3981 candidates.
Taking the conservative approach is also relevant for our nova search. For this case, besides the existence of  false matches, there are two more reasons to consider all candidates (with and without matches in the reference image). Firstly,  the reference image used for cross-matching has been constructed (via co-addition) using a subset of observations from the same data set used for searching the novae and therefore it can be contaminated by some novae. Secondly,  there is no sharp cut-off for the quiescence magnitudes of novae; for example, the quiescence magnitudes of novae in M31 measured by \citet{Williams-2014} reach values of $m_{R}\sim 22$.

\section{A systematic search for novae in the iPTF M31 observations}
\label{nova_search}

With the set of unique candidates for variable and transient sources obtained via \texttt{WAVDETECT} (Sect.~\ref{wavdetect}), it becomes feasible to search for any class of objects of interest. Hereupon we simply need to resort to the routine procedure of constructing lightcurves and making the selection based on the expected behavior of the objects sought. With an efficient algorithm, the few thousand candidates obtained in Sect.~\ref{wavdetect} can be reduced to a set of a few, optimized for the class of objects of interest. In particular, we make a systematic search for novae among the 3981 variable and transient source candidates found above.

Once the locations of the different candidate sources are identified, we proceed to construct their lightcurves using the whole set of {\em difference images} per candidate. 
We perform forced photometry at these positions, i.e., we measure the fluxes of the candidates at all epochs irrepsective of whether or not they are detected in a given epoch by the iPTF DI pipeline. As we are measuring fluxes on difference images, which are relatively sparsely populated and stellar blending is not expected, we have opted for simple aperture photometry. This is executed using the PHOT package of DAOPHOT \citep{Stetson} and with curve of growth correction using DAOGROW \citep{Stetson-1990}. The resulting flux measurements thus represent the differential fluxes of the variable and transient source candidates. Note that the measured fluxes are in detector units, i.e., DN.

\begin{figure*}
 \centering
\includegraphics[width=88mm]{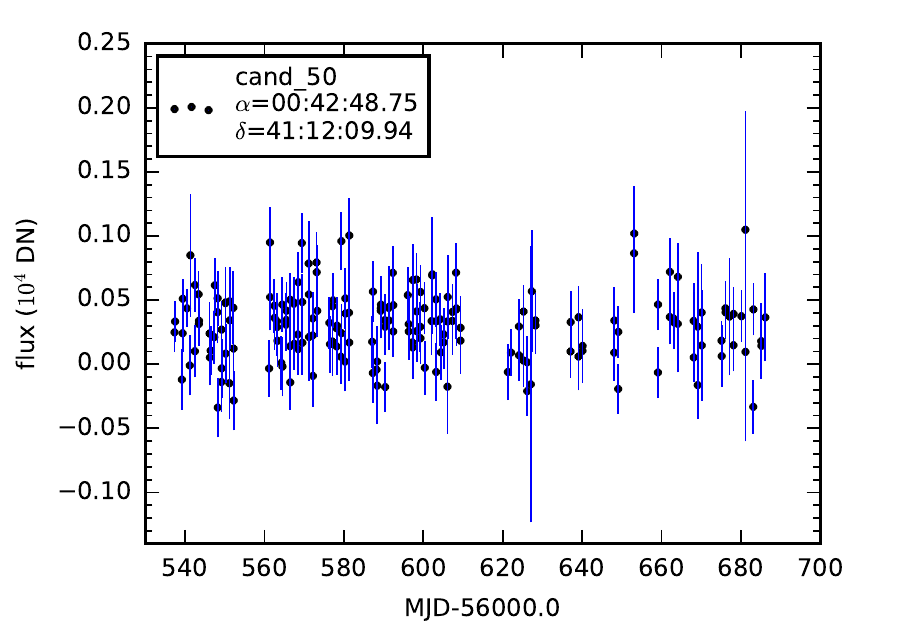} \hfill \includegraphics[width=88mm]{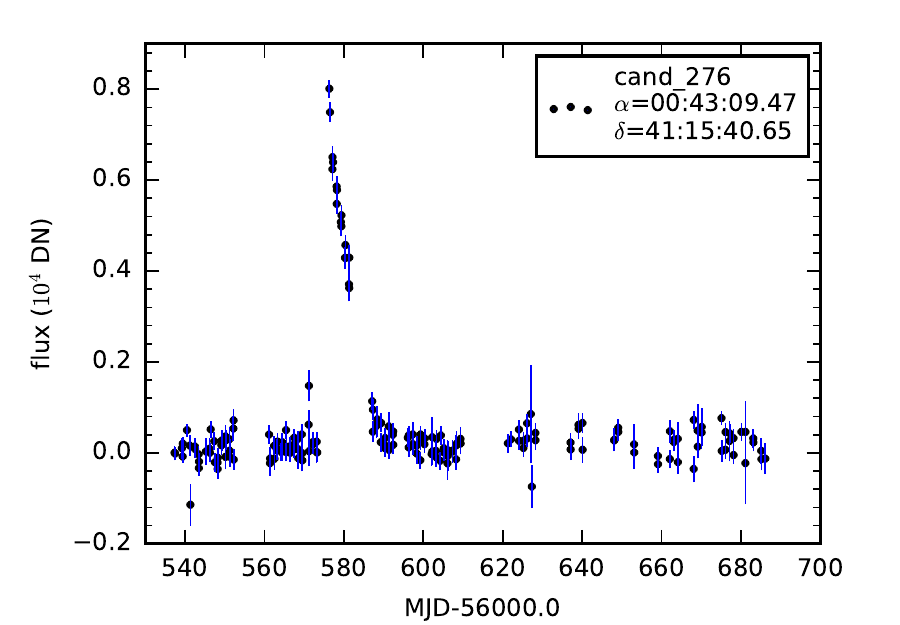}
\vspace{10mm} 
\includegraphics[width=88mm]{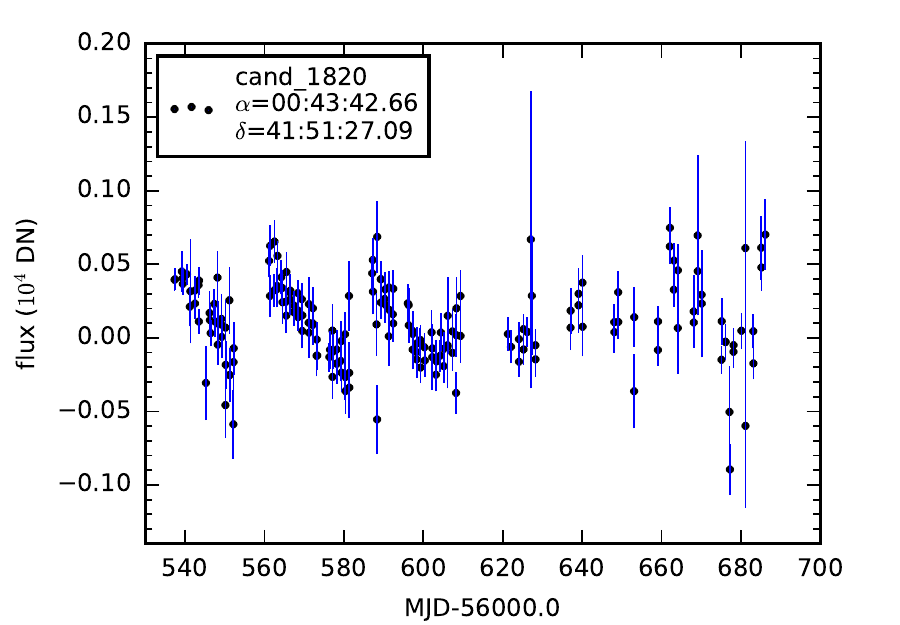} \hfill \includegraphics[width=88mm]{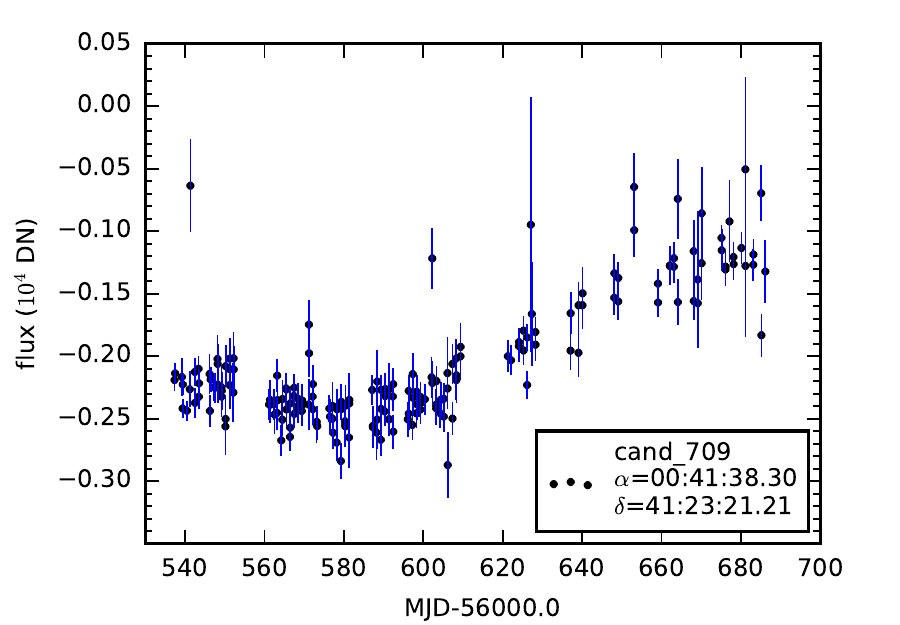}
\caption{Examples of lightcurves of four candidates identified by WAVDETECT, constructed via forced aperture photometry on the difference images. The lightcurve in the top left panel is that of an artifact resulting from imperfect PSF-matching in the DI, while the one in the right panel is that of a nova (cf.~Sect.~\ref{results}). The two lightcurves in the bottom are that of genuine variable sources. For the source in the bottom right panel, all the measured fluxes have negative values. This occurs because the reference image used in the difference imaging is not an unbiased average of the random phases of the lightcurve.}
\label{fig:eg_lcs}
\end{figure*}

Four examples of the resulting lightcurves are shown in Fig.~\ref{fig:eg_lcs}. 
As can be seen in the figure, three of the lightcurves exhibit clear and distinct features of variability; in fact, one is the lightcurve of a nova (Fig.~\ref{fig:eg_lcs} top right panel), as we will confirm in Sect.~\ref{results}. The remaining fourth lightcurve is almost flat and featureless. From visual inspection, we verified this latter candidate to be an artifact resulting from imperfect PSF-matching in the DI.
Further, some outlier data points can be clearly seen, for example in the bottom right panel of Fig.~\ref{fig:eg_lcs}. These spurious flux measurements also arise from low-quality difference images.

The lightcurves of all the variable and transient source candidates found by \texttt{WAVDETECT} are then filtered for novae, as described in the following subsection.

\subsection{Lightcurve filtering for novae}\label{novae} 

We expect a nova lightcurve to exhibit a contiguous or pseudo-contiguous (see below) section of excess brightness marking the eruption, while being consistent with the quiescent background before and probably after the eruption. We build our lightcurve-filtering algorithm on this premise. 
The steps implemented are the following:
\begin{enumerate}
\item {A baseline is determined iteratively by computing the mean of the 2-sigma clipped lightcurve using all the data points. 
The clipping is carried out both above and below the mean, using the rms deviation about the mean.
In case there are fewer than 10 points remaining before convergence of the clipping process (for example this might happen in periodic variables), we use all the data points in computing the mean. The final mean and the rms deviation are then used as the baseline and $\sigma$, respectively, in the following steps.} 
\item {All the data points in the lightcurve that are more than $4\sigma$ above the baseline are identified. Let us call this set of points $S_{\rm p}$ and the number of points in it $n_{\rm p}$. It does not matter here whether these points are contiguous or not.}
\item {All the data points in the lightcurve that are more than $4\sigma$ below the baseline are also identified. Let us call this set $S_{\rm n}$ and its number of points $n_{\rm n}$.}
\item {We require that the nova candidate lightcurves have $n_{\rm p}\ge 4$ and $n_{\rm n}\le {\rm max}\{3, 0.30\times n_{\rm p}\}$. The latter condition helps in filtering out periodic variables.}
\item {We take the observation times (i.e., the modified Julian date MJD) corresponding to the first and last points of $S_{\rm p}$, $t_{\rm min}$ and $t_{\rm max}$, respectively, to bracket the main outburst part of the nova. Let $n$ denote the {\em total} number of data points in the lightcurve between $t_{\rm min}$ and $t_{\rm max}$. If $n_{\rm p}\ge 0.70\times n$, then step 7 is executed, else step 6. The factor 0.7 is chosen somewhat arbitrarily through experimentation.}
\item {If $n_{\rm p}< 0.70\times n$, the following steps are implemented to address the possible presence of outlier flux measurements (see above) contaminating potential nova candidate lightcurves.

\begin{enumerate}[listparindent=2.5em]
\item {

In principle, for uniform temporal coverage, we could do the following. We can take the observation MJDs corresponding to the $S_{\rm p}$ points and compute an array of interval in time between consecutive $S_{\rm p}$ points. The spurious data points will have large values and will be outliers in such an array. The iPTF coverage however is not uniform, having gaps that could be even greater than a week. Nevertheless, the above method can still be implemented by working simply with array indices, as detailed below.

We first make an array containing all the data points of the lightcurve, sorted in increasing order of the observation MJDs.
Using the indices ($i$) from this array corresponding to the $S_{\rm p}$ points, we create a new array, let us call it {\em index-interval array}, containing their forward differences (i.e., $i_{\rm k+1}-i_{\rm k}$). 
We can get rid of outlier data points in $S_{\rm p}$ by partitioning the index-interval array using some statistics of the array itself. We choose to use the median $M$ and the rms deviation about the median $\sigma_{\rm M}$, determined from the 3-sigma clipped index-interval array. The clipping is performed iteratively {\em above} the mean of the array, using the rms deviation about the mean. If convergence is not achieved by the time the number of elements in the clipped array reduces to 80\% of the original number, we proceed with the immediately preceding clipped index-interval array. Using the so-determined $M$ and $\sigma_{\rm M}$, we partition the original index-interval array into segments $L_i$ at positions where the array value is $>{\rm max}\{2, M+1.5\sigma_{\rm M}\}$.
}
\item {We take the longest segment max($L_i$) as the new $S_{\rm p}$, with $n_{\rm p}$ denoting its number of elements. We then determine $t_{\rm min}$, $t_{\rm max}$ and $n$ as in Step 5. If $n_{\rm p}\ge 0.70\times n$, step 7 is executed, else the candidate is discarded.}
 \end{enumerate}
\item{Finally, we apply a rise-time filter to the lightcurves that have passed through step 5 to eliminate long-period variable stars. In a nova, the initial rise takes at most 3 days; this is followed by a pre-maximum halt that lasts for a few hours to few days, and the final $\sim 2$~mag rise to the peak takes some days to weeks \citep{Warner-1995}. Accordingly, we require the rise-time of the candidate lightcurve to be less than 2 weeks. The rise-time is defined as $t_{\rm peak}-t_{\rm min}$, where $t_{\rm peak}$ is determined by fitting a broken power law to $S_{\rm p}$ with one breakpoint corresponding to the peak.}
}
\end{enumerate}

\begin{sidewaystable*}
\renewcommand*{\arraystretch}{1.5}
\captionsetup{justification=centering}
\caption{iPTF M31 nova sample for the period between 09/2013 and 01/2014.}
\label{tab:recovery}
\centering
\begin{tabularx}{\textwidth}{l l l l l c c X}
\hline
ID	&RA\tablefootmark{a}	&DEC\tablefootmark{a}	&$m_{R, {\rm peak}}$	&$t_2$(days)\tablefootmark{b,c}	&Known nova name	&Reference	&Comments\tablefootmark{d}\\
\hline
89	&0:43:00.47	&41:12:36.71	&15.8	&$10.3\pm0.3$	&M31N 2013-09c	     &1	&\\
234	&0:43:13.55	&41:14:47.69	&16.4	&$11.2\pm2.6$	&M31N 2014-01a	     &2	 	&spectroscopically confirmed; type Fe II\\
243	&0:42:49.87	&41:14:57.81	&18.0	&$28.0\pm4.9$ -- $44.6\pm4.8$	&M31N 2013-12a	     &3	 	&spectroscopically confirmed; type narrow-lined He/N\\
276	&0:43:09.47	&41:15:40.65	&17.3	&$10.4\pm0.4$	&M31N 2013-10c	     &1	&\\
319	&0:43:04.90	&41:16:31.13	&14.8	&$6.4\pm0.3$	&M31N 2013-10h	     &4	 	&spectroscopically confirmed; type Fe II\\
355	&0:42:46.52	&41:17:00.84	&16.7	&$24.2\pm3.5$ -- $83.0\pm7.2$	&M31N 2013-10b	     &5	&\\
376	&0:42:42.35	&41:17:18.86	&16.0	&$7.0\pm0.2$	&M31N 2013-10e	     &1	&\\
427	&0:43:24.21	&41:18:19.49	&16.4	&$19.6\pm0.5$ -- $21.6\pm0.9$	&M31N 2013-10a	     &6	&spectroscopically confirmed; type Fe II\\
430	&0:42:51.66	&41:18:14.71	&16.7	&$13.1\pm1.1$ -- $16.1\pm1.4$	&M31N 2013-12b	     &7	 	&spectroscopically confirmed; type hybrid\\
473	&0:43:14.93	&41:19:13.35	&16.5	&$12.1\pm0.6$	&M31N 2013-09a	     &8	 	&\\
601	&0:43:24.90	&41:21:22.38	&17.4	&$43.8\pm1.8$ -- $47.1\pm1.2$	&M31N 2013-10g	     &9	 	&spectroscopically confirmed; type Fe II\\
981	&0:42:23.55	&41:29:12.68	&18.9	&--	&M31N 2014-01b	     &10	 	&spectroscopically confirmed\\
1626	&0:41:47.02	&41:44:11.59	&19.1	&$28.6\pm8.6$ -- $110.8\pm4.7$	&--	     &this work			&possible nova\\
1687	&0:43:50.44	&41:46:12.36	&19.3	&--	&--	     &this work			& nova candidate (faint, fast)\\
3182	&0:43:15.56	&42:13:51.34	&19.5	&$3.0\pm1.3$ -- $8.9\pm2.3$	&--	     &this work			& nova candidate\\
\hline
--	&0:42:34.94	&41:14:56.30		&--	&--	&M31N 2014-01c	     &10	 	&spectroscopically confirmed; type Fe II. Erupted at the end of the iPTF season analyzed here\\
\hline
\end{tabularx}
\tablefoot{
\tablefoottext{a}{ICRS coordinate system (J2000.0).}
\tablefoottext{b}{When there are fluctuations in the lightcurve, two values of $t_2$ are given corresponding, respectively, to the first and last times the lightcurve falls by 2~mags from peak.}
\tablefoottext{c}{The errors on the $t_{2}$ values are estimated by a bootstrapping algorithm (see text).}
\tablefoottext{d}{Information about the spectroscopic types for the novae here is included whenever available (for details, see for example, \citealt{Williams-1992}, \citealt{Shafter-2011}).}
}
\tablebib{
(1)~\citet{nova2+3+5+6}; (2)~\citet{nova11}; (3)~\citet{nova9b}; (4)~\citet{nova8b}; (5)~\citet{Hornoch-2013}; (6)~\citet{Bigley-2013}; (7)~\citet{nova10b}; (8)~\citet{nova1}; (9)~\citet{nova7b}; (10)~\citet{nova12b}.
}

\end{sidewaystable*}

After passing the 3981 candidate lightcurves through the above selection algorithm, we obtain a set of 15 nova candidates, which are listed in Table~\ref{tab:recovery}. Estimates of the lightcurve peak value (obtained from Step~7 above) expressed in calibrated magnitude $m_{R, {\rm peak}}$ and decline times $t_2$ are also included in the table. These $t_2$ values have been obtained by linear interpolation of the data points; when fluctuations are present in the lightcurves, we give values of both the {\em first} ($t_{2i}$) and {\em last} ($t_{2f}$) times the lightcurve falls by 2~mags from peak (or in linear scale when the flux has declined by a factor of $10^{2/2.5}$ from the peak value). We determine the accuracy of the $t_2$ estimates by employing a bootstrapping algorithm, whereby we resample a given lightcurve 1000 times by randomly selecting half of the data points and recomputing the $t_2$ values each time. We then obtain the uncertainty by computing the standard deviation of the 1000 $t_2$ values\footnote{Due to the correlated nature of the time-series data, the bootstrapping algorithm does not strictly provide a formal one-sigma error bar. The quoted uncertainties should rather be interpreted as a rough indication of the precision of the $t_2$ estimates.}.
It is to be noted that for candidates 1626 and 355, the flux corresponding to 2~mags from the peak is within $1\sigma$ of the baseline already, and therefore for these two candidates, a better estimate of their $t_2$ could be provided by $t_{2i}$. Furthermore, for candidates 1687 and 981, their $t_2$ times cannot be estimated as their lightcurves have not yet declined by 2~mags from the peaks in the period analyzed here. 
Of course, slow novae, whose duration of eruptions is comparable to or much longer than the baseline of the observations presented here (five months) will be missed in our sample.

\section{Results}\label{results}

The lightcurves of the final 15 nova candidates are shown in Fig.~\ref{fig:nova-lcs}. The identification number (ID) of the candidates from Table~\ref{tab:recovery} are included in the legends on the plots. 
From the M31 optical nova catalog maintained at MPE\footnote{\label{MPE}{\url{http://www.mpe.mpg.de/~m31novae/opt/m31/M31_table.html}}} \citep{Pietsch-2007}, we are able to verify that 12 of them are in fact confirmed novae. 
The names of these candidates in the MPE catalog are also shown in Table~\ref{tab:recovery}.

During the five-month period between September 2013 and January 2014, when these data were taken, 13 novae (occurring in the field examined here) have been registered at the MPE catalog by different authors based on observations with different telescope facilities. We list the 13 novae in Table~\ref{tab:recovery}. We have recovered all except one, i.e. M31N~2014-01c. This nova occurred close to the center of M31 at the end of January 2014, which is near the end of the season analyzed here. The iPTF DI pipeline detected this nova only 5 times when it erupted. This value is, however, insufficient for its detection in the densely crowded bulge (cf.~Fig.~\ref{fig:hits}) by \texttt{WAVDETECT}, which requires at least 7 hits (see  Sect.~\ref{wave_counts}). With data extending beyond January 2014, this nova would most likely have been detected by \texttt{WAVDETECT}. 
For the remaining three nova candidates in our sample (1626, 1687 and 3182), candidate 1626 appears clearly a nova, but for candidates 1687 and 3182, we cannot firmly establish their nova nature from the present lightcurves since their variability is detected at the end of the observation period analyzed here. Still, candidate 1687 could be a fast faint nova similar to  M31N~2014-01b (ID 981). In this work we have thus found one new nova and two nova candidates in M31 analyzing the five-month-long data set.

From the discussion above, the MPE nova catalog appears to be comparable with the iPTF nova sample obtained here. The former is mainly a result of work  of the many amateur astronomers who made the discoveries of the individual novae.     
However, data from surveys are better-suited for statistical studies of the nova population, since surveys provide the advantage of uniform coverage in various aspects of the observations, for example detectors used, that allows a robust computation of the sample completeness  (see also Sect.~\ref{discuss}).
The fact that the whole procedure involved in this search for novae is automated, makes it possible to quantitatively characterize any incompleteness in the sample. This is left for future work, following the analysis of the much larger full iPTF data set of M31.

\section{Discussion and summary}\label{discuss}

We have presented a new and efficient method for tackling the menace of artifacts contaminating the genuine variable and transient  sources detected by automated data pipelines of time-domain surveys. We have used a five-month-long string of iPTF observations of M31 covering its crowded bulge, with $\sim 2.5\cdot 10^5$ raw detections by the iPTF DI pipeline, to illustrate the methodology. 
Using all the pipeline detections, we created an image, termed the ``hits-image'', each pixel of which records the number of times a signal from the given position was registered on the difference images (Sect.~\ref{space_time}, Figs.~\ref{fig:cluster}, \ref{fig:hits}). 
We have shown that about half of the artifacts form a locally uniform (nearly) Poissonian background in the hits image (Fig.~\ref{fig:poisson}), with the remaining half being associated with bright stars. The latter, as well as genuinely  variable and transient sources appear as clusters of detections, much like in an image produced by a grazing incidence X-ray telescope (cf.~Fig.~\ref{fig:cluster} lower panel). This analogy with an X-ray image allows us to import various well-established tools in X-ray astronomy without requiring any modification.  
Multi-scale wavelet analysis provides an efficient and convenient means to detect structures/sources in such images. We have thus chosen the popular wavelet-based tool \texttt{WAVDETECT} from the {\it Chandra}'s \texttt{CIAO} package \citep[see][]{Freeman-2002} for the identification of the clusters of detections in the hits image. 
Running \texttt{WAVDETECT} on it, we obtained the unique candidates for variable and transient sources for the whole observation season, numbering $\sim 4\cdot10^{3}$. 

Cross-matching the \texttt{WAVDETECT} ``sources'' with the source-catalog of a deep reference image of the same M31 field, we found $\sim 90\%$ of them to have counterpearts in the reference image; in fact almost all sources in the reference image brighter than $m_{R}=18$ have been detected multiple times in the difference images (Fig.~\ref{fig:reference_mag}). This may be due to some of the bright sources being genuinely variable, and others occasionally producing artifacts due to imperfect PSF-matching in the DI pipeline. Only a mere $\sim 450$ candidates remained without counterparts on the reference image. These sources are candidates for transient sources (different from variable sources in this context). Thus we achieved an almost three orders of magnitude reduction from the initial number of raw DI pipeline detections of $\sim 2.5\cdot 10^{5}$.

The method presented here is, in principle, well-suited for archival, but not real-time time-domain data analysis. However, for the modern time-domain surveys (for example iPTF and the upcoming LSST), cadences for particular fields could range from hours to even minutes per observation, such that the total number of visits during the night can sum up to more than a few. 
In such cases, our method can be straightforwardly implemented to search  for variable and transient sources near real-time.

The prospects of this method are attractive when it comes to investigating the {\em populations} of variable and transient objects. It can successfully probe the variable and transient sources even in the regions severely affected by low quality of the DI, for example in the bulge of M31 or in the Galactic plane, which could not practically be achieved earlier in large-scale surveys. The method can be implemented within the automated DI pipeline of the surveys, which will then directly yield outputs ready for astrophysical studies.

We have illustrated the use of the proposed method by applying it to  a systematic search for novae in the iPTF data collected during the period from September 2013 to January 2014. To this end, we developed a lightcurve-based filtering procedure to search for novae among the candidates identified by \texttt{WAVDETECT}.
We found 15 nova candidates, of which 12 are known novae listed in the MPE nova catalog \citep{Pietsch-2007}. Of the remaining 3 candidates, we consider one (candidate 1626, Fig.~\ref{fig:nova-lcs}) as a very likely nova based on the shape of its lightcurve and two more (candidates 1687 and 3182) occurred near the end of the analysed data and need further investigation.  We have recovered all novae that were reported to occur during the period of these observations in the MPE catalog, except for  M31N~2014-01c that erupted at the very end of the iPTF observations. Four (including all three newly discovered candidates) of the novae detected by iPTF (IDs 981, 1626, 1687 and 3182, Fig.~\ref{fig:nova-lcs}) occurred far away from the bulge region and thus clearly in the disc of the galaxy. The remaining 11 appear to be concentrated around the bulge region. We do not make any inferences regarding the specific nova rates in the disc and bulge of the galaxy. This will be addressed in the future work based on the $\gtrsim 10$-fold larger full iPTF data set.

The fact that all the procedures involved in the  novae search --- from DI to final lightcurve filtering --- are automated will allow one to compute the incompleteness of the sample, as will be done in a future work based on the   full iPTF data set. However, we can  make a crude estimate of the completeness of the present sample. From \citet{Soraisam-2016}, the approximate total nova rate for M31 galaxy, counting fast novae,  is $\approx 106$/yr. The iPTF M31 field analyzed here encloses roughly 40\% of the total stellar mass of M31, and therefore we expect $\sim 17$ novae to occur during the five-month period of the iPTF M31 observations used here. In our analysis, we have detected  13 (15, if counting the two candidates) novae, which implies a completeness of the sample of $\sim 76\%$ ($\sim 88\%$). With the high cadence of the iPTF survey, we anticipate the future  iPTF  M31 nova sample to be particularly useful to constrain population of  novae   with short decline times. These fast novae are predicted by theories \citep[e.g.,][]{Yaron} to occur on massive white dwarfs (WDs) and therefore are imperative for determining the WD mass distribution in novae toward the interesting massive end \citep[see][]{Soraisam-2016}.

These results thus demonstrate that the method we have presented in this paper provides an efficient, yet simple, way to analyze the outputs of time-domain data pipelines. By suitably modifying the lightcurve selection algorithm, the method can be easily applied to any survey of any field for probing a wide range of transient sources, including periodic variables such as Cepheids.

\section*{Acknowledgments}
MDS is grateful to Niels Oppermann (CITA) for helpful discussions on the Bayesian treatment of obtaining the probability distribution of false matches.
MDS thanks the California Institute of Technology, where a portion of this work was completed, for its hospitality. MDS was supported, in part, by the GROWTH internship program funded by the
National Science Foundation under Grant No 1545949. AWS is grateful to the NSF for financial support through grant AST-1009566.
MG acknowledges partial support by Russian Scientific Foundation (RSF), project 14-22-00271.

\bibliographystyle{aa}
\bibliography{ref}

\begin{figure*}
\centering
\includegraphics[width=88mm]{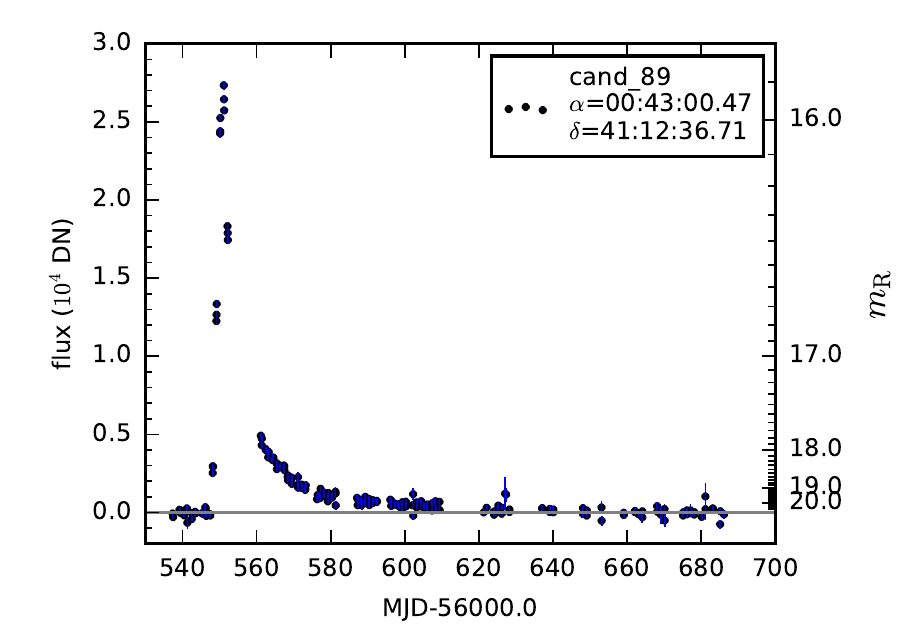} \hfill \includegraphics[width=88mm]{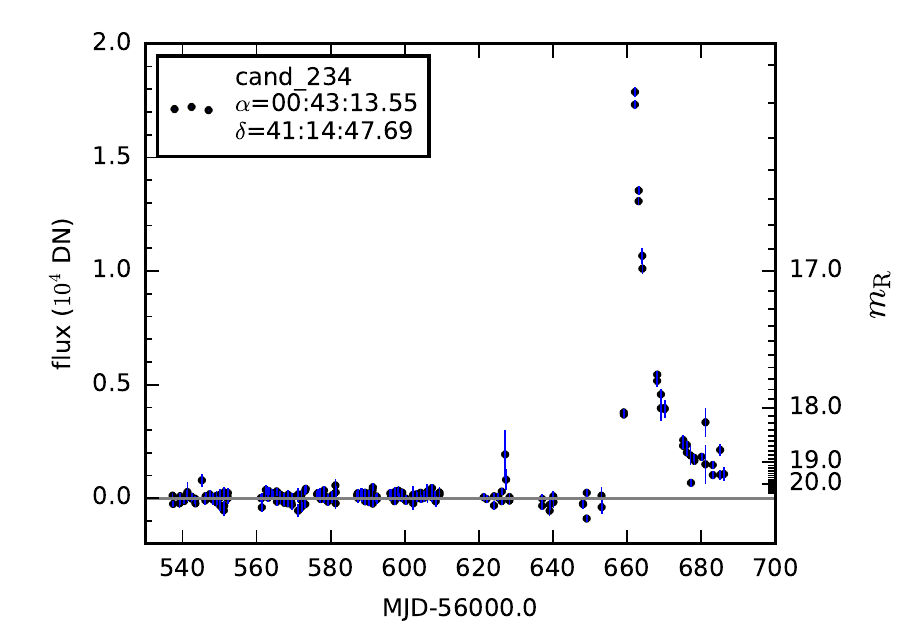}
\vspace{10mm}
\includegraphics[width=88mm]{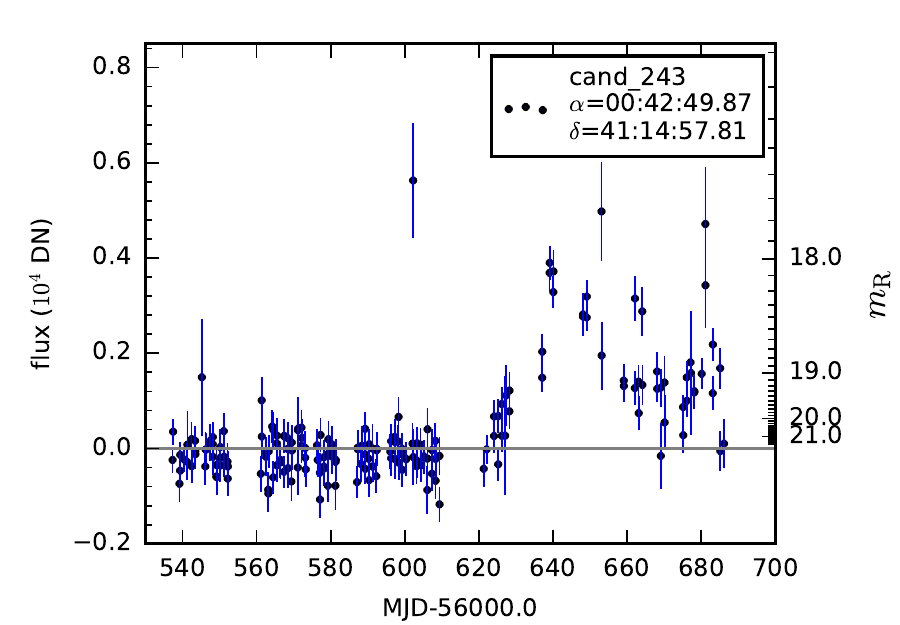} \hfill \includegraphics[width=88mm]{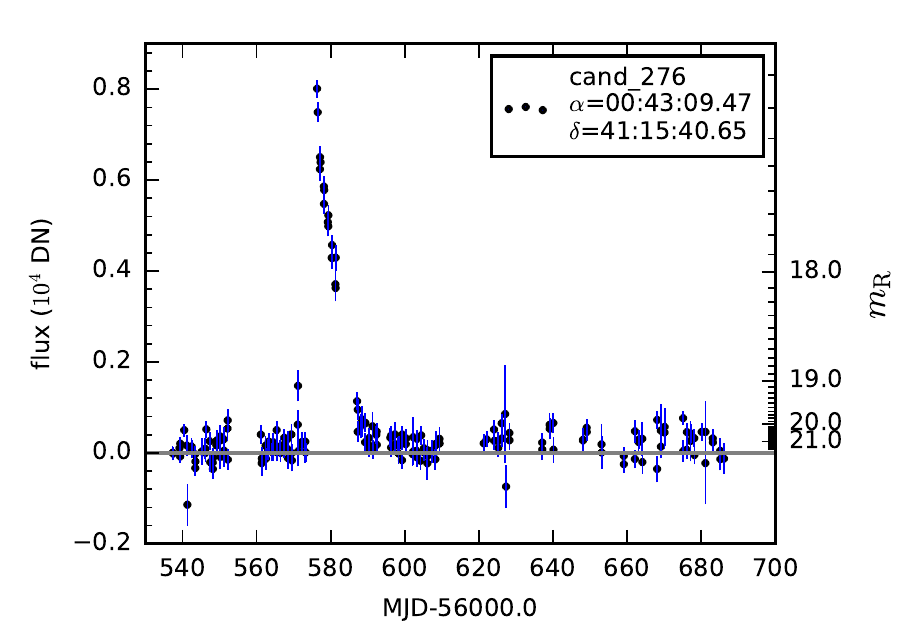}
\vspace{10mm}
\includegraphics[width=88mm]{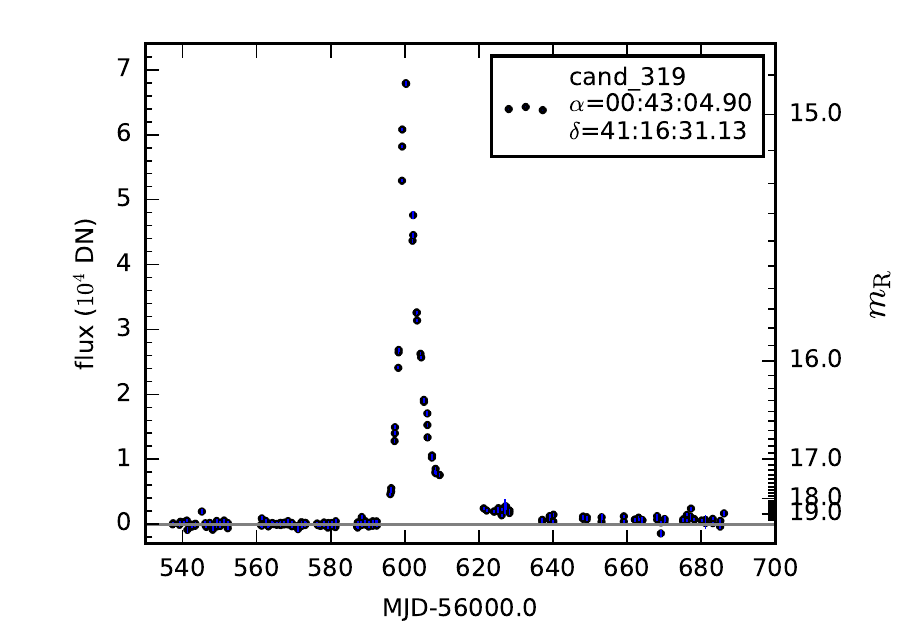} \hfill \includegraphics[width=88mm]{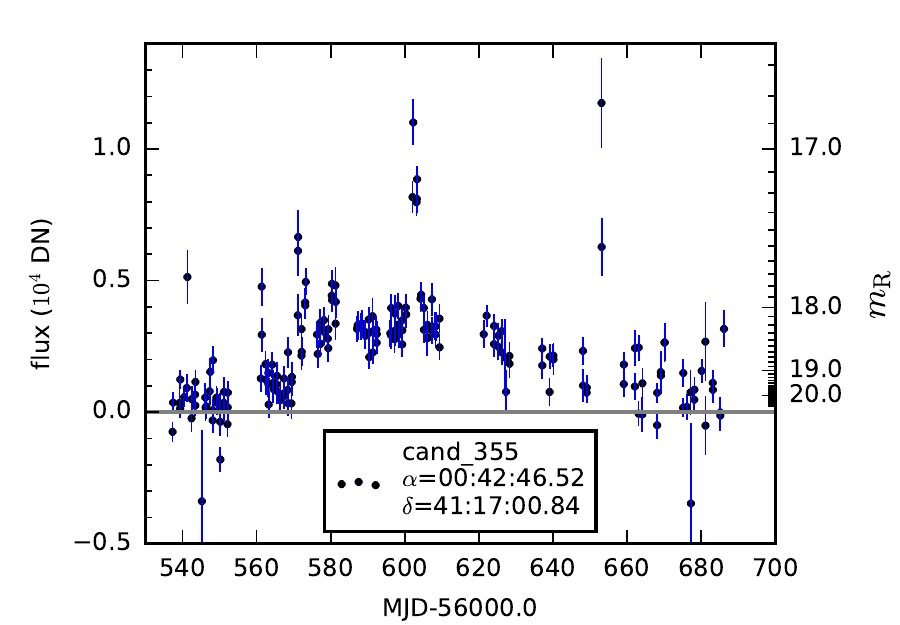}
\caption{Lightcurves of the nova candidates obtained from the nova selection algorithm (see Sect.~\ref{novae}). The vertical axis on the right side shows the calibrated $R$-band magnitudes corresponding to the flux values on the left vertical axis.} 
\label{fig:nova-lcs}
\end{figure*}

\setcounter{figure}{7}
\begin{figure*}
\centering
\includegraphics[width=88mm]{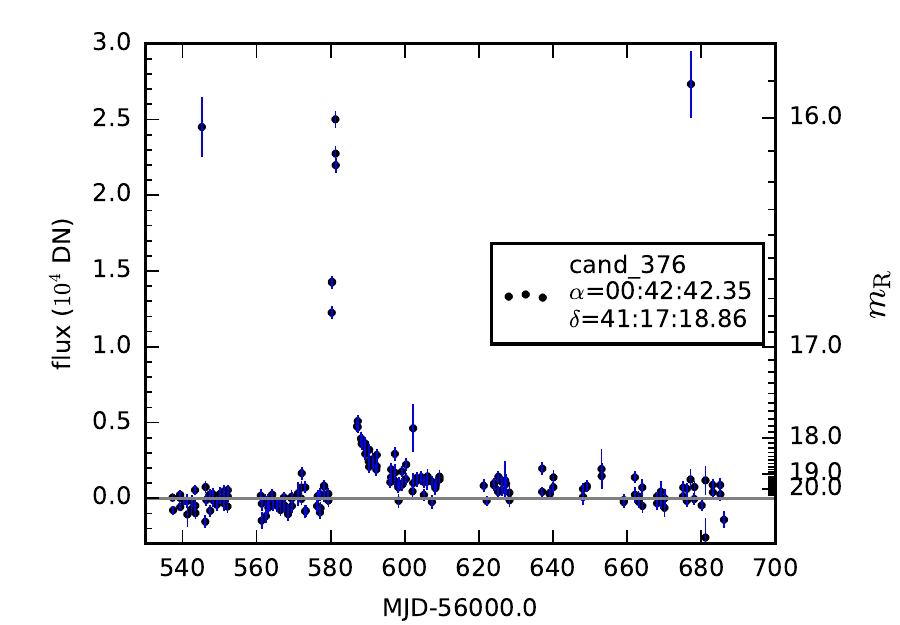} \hfill \includegraphics[width=88mm]{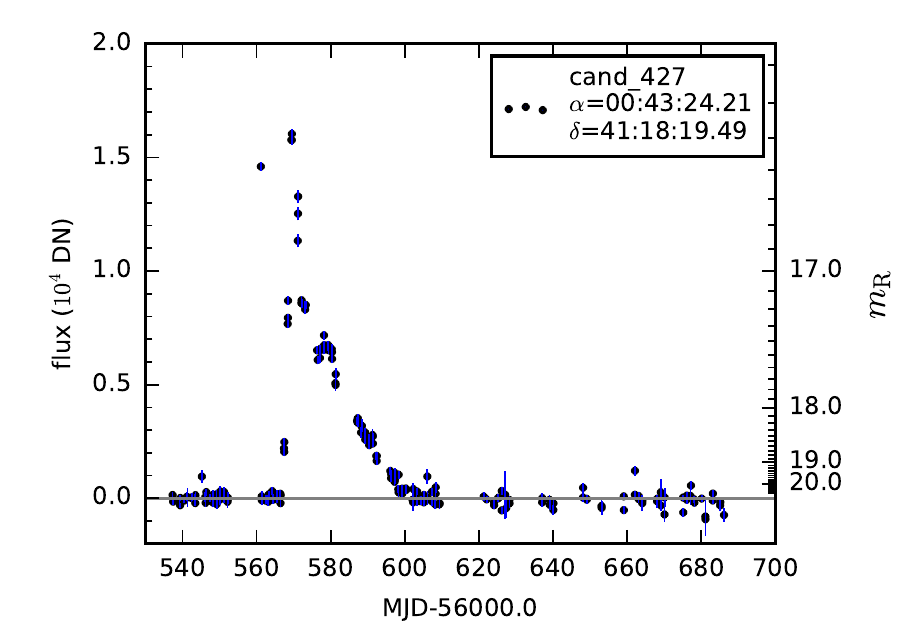}
\vspace{10mm}
\includegraphics[width=88mm]{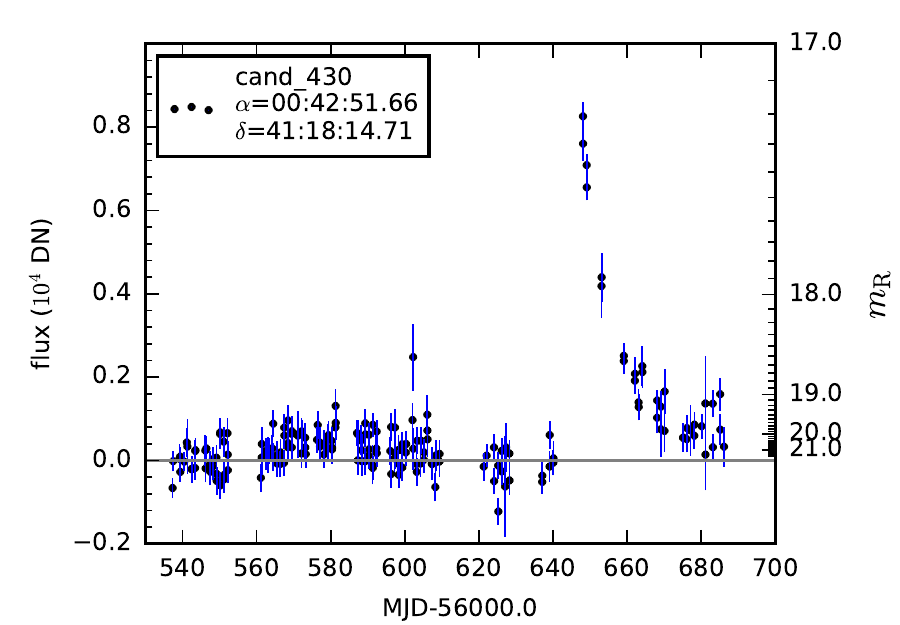} \hfill \includegraphics[width=88mm]{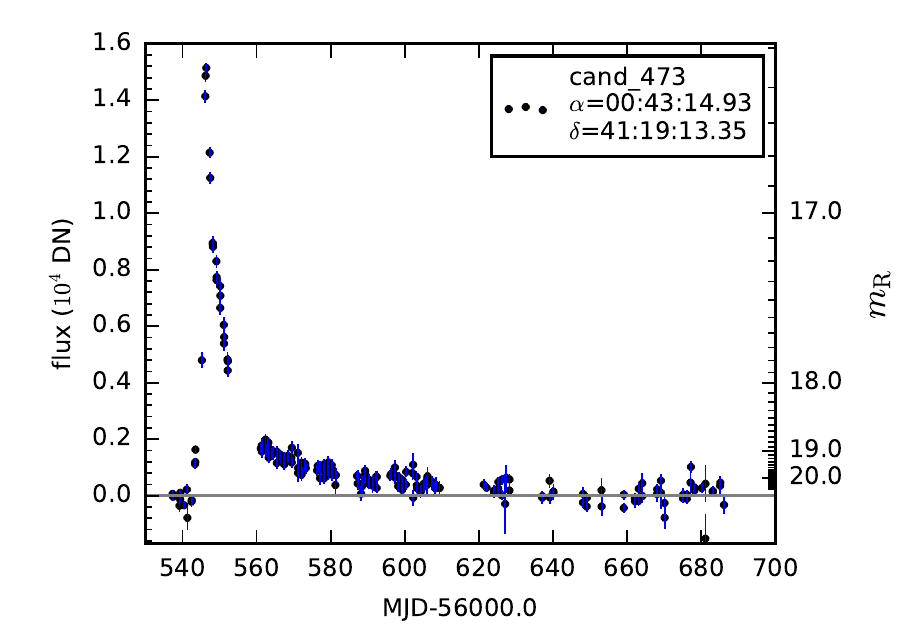}
\vspace{10mm}
\includegraphics[width=88mm]{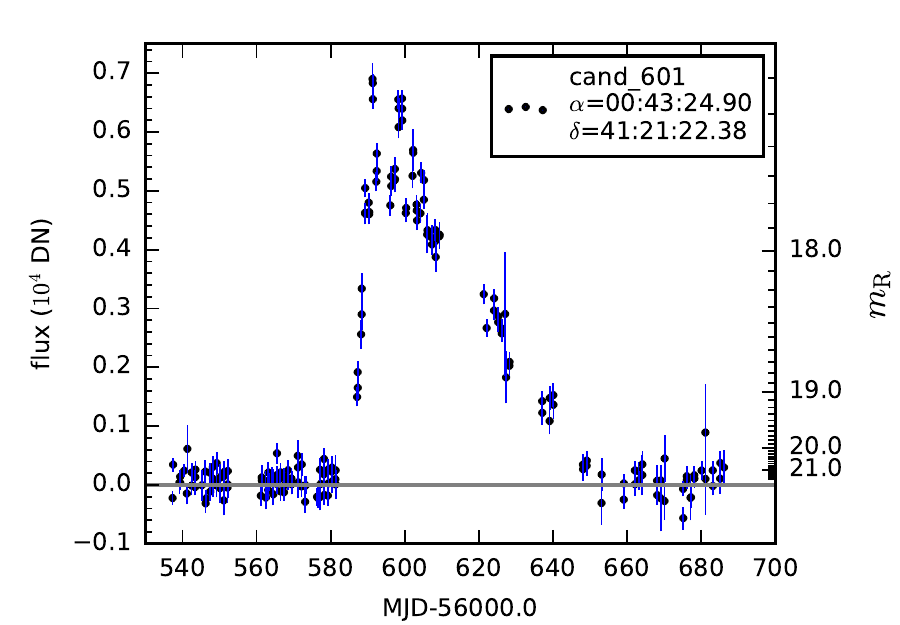} \hfill \includegraphics[width=88mm]{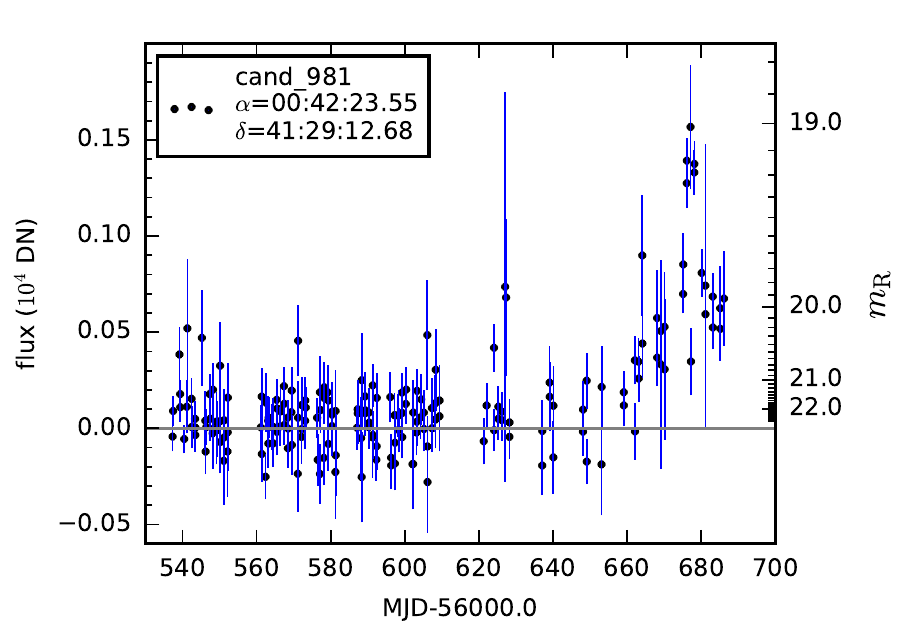}
\caption{continued.}
\end{figure*}

\setcounter{figure}{7}
\begin{figure*}
\centering
\includegraphics[width=88mm]{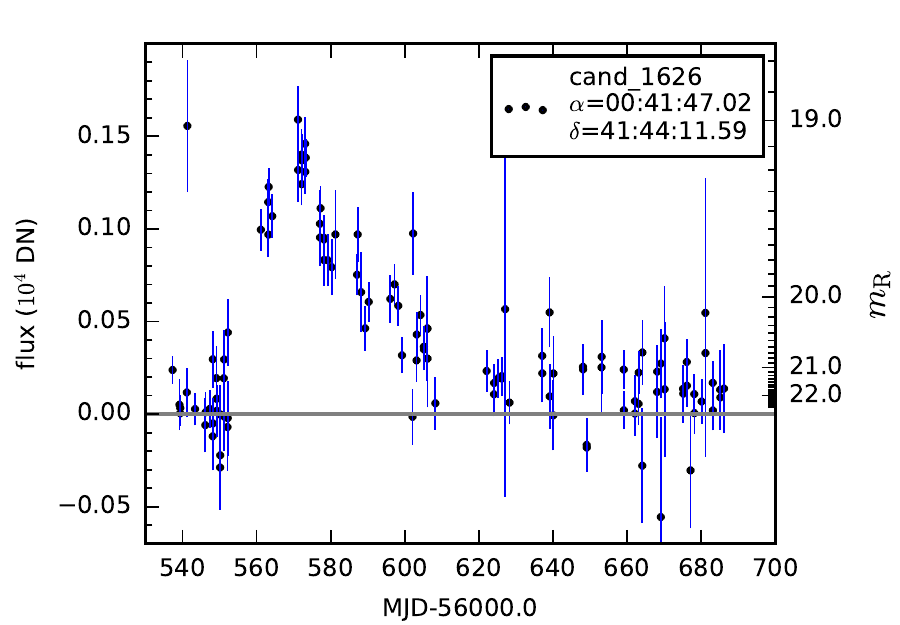} \hfill \includegraphics[width=88mm]{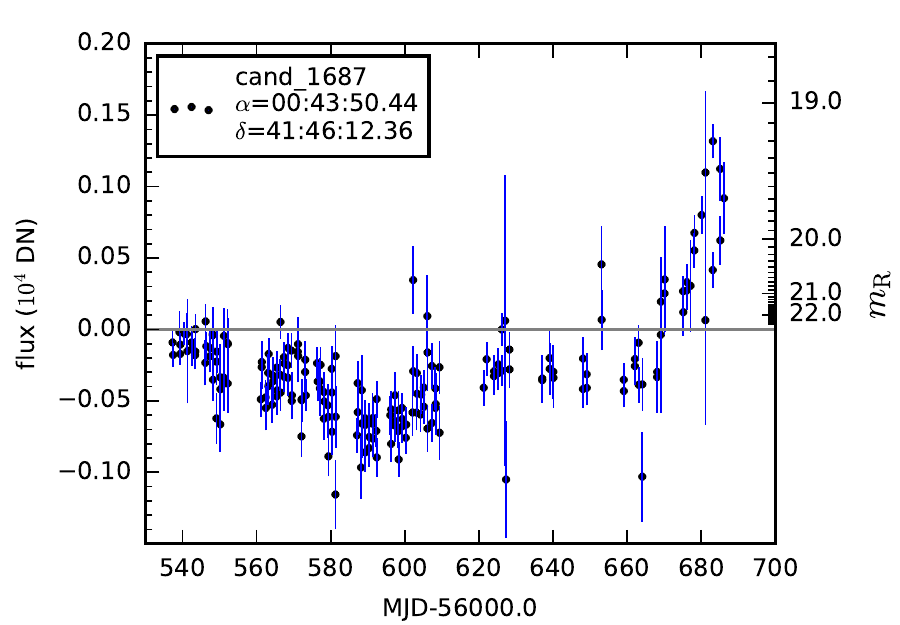}
\vspace{10mm}
\includegraphics[width=88mm]{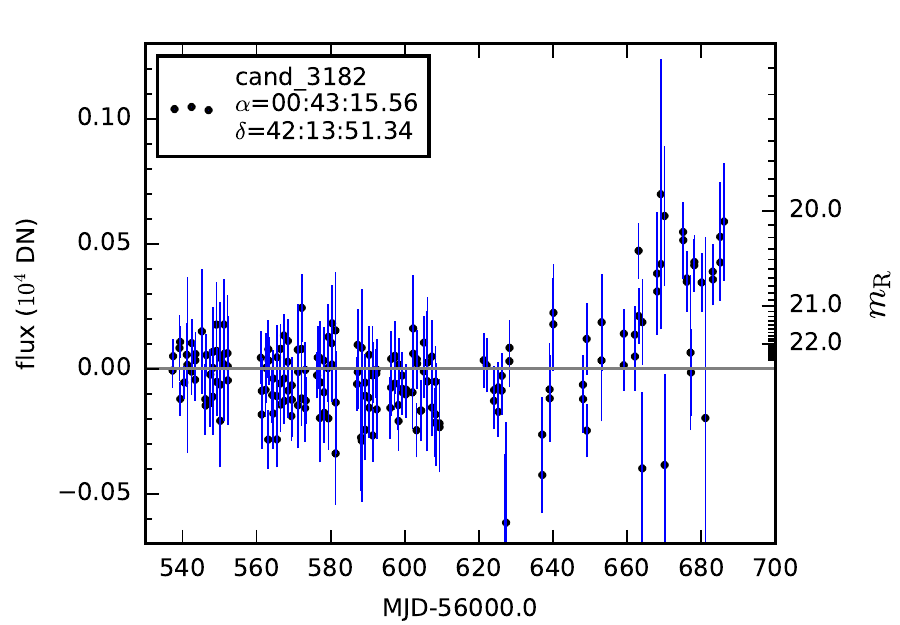}
\caption{continued.}
\end{figure*}

\appendix

\section{Probability distribution of false matches}\label{rand_match}

The unknown numbers of true ($N_{\rm t}$) and false ($N_{\rm f}$) matches are related to the known number of matched \texttt{WAVDETECT} sources $N_{\rm m}$ by the obvious relation
\begin{equation}
N_{\rm m}=N_{\rm t}+N_{\rm f}.
\label{eq:false_match1}
\end{equation}
Assuming a uniform distribution of the sources in the reference image across the region of interest, the expected number of reference image sources occurring randomly within the matching area is given by 
\begin{equation}
\mu(N_{\rm t})=\pi r^{2}(N_{\rm r}-N_{\rm t})/A,
\end{equation}
where $A$ is the known area under study, $r$ is the known matching radius and $N_{\rm r}$ is the known number of reference image sources in this area. The probability of having one or more random matches for a given source is then 
\begin{equation}
p=p(N_{\rm t})=1-e^{-\mu(N_{\rm t})}.
\label{small_p}
\end{equation}

If $N_{\rm t}$ is known, the probability distribution of false matches for the $N_{\rm w}-N_{\rm t}$ sources ($N_{\rm w}$ is the known number of total \texttt{WAVDETECT} sources), which is binomial, can be written as 
\begin{equation}
P(N_{\rm f}|N_{\rm t})=\binom{N_{\rm w}-N_{\rm t}}{N_{\rm f}}[p(N_{\rm t})]^{N_{\rm f}}[1-p(N_{\rm t})]^{N_{\rm w}-N_{\rm t}-N_{\rm f}},
\label{eq:binomial}
\end{equation}
where $p(N_{\rm t})$ is given by Eq.~\eqref{small_p}.

We then relate the above probability distribution to the one that we are interested in, that is $P(N_{\rm f}|N_{\rm m})$, by using Bayes' Theorem and marginalizing over various quantities, as
\begin{align}
P(N_{\rm f}|N_{\rm m})&=\sum^{\infty}_{N_{\rm t}=0}P(N_{\rm f},N_{\rm t}|N_{\rm m})\\
                      &=\sum^{\infty}_{N_{\rm t}=0}P(N_{\rm f}|N_{\rm t},N_{\rm m})P(N_{\rm t}|N_{\rm m})\\
                      &=\sum^{\infty}_{N_{\rm t}=0}P(N_{\rm f}|N_{\rm t},N_{\rm m})\frac{P(N_{\rm m}|N_{\rm t})P(N_{\rm t})}{P(N_{\rm m})}.
\end{align}
Here, $P(N_{\rm m})$ is a normalization constant. Assuming the prior $P(N_{\rm t})$ to be uniform, we introduce the constant $C=P(N_{\rm t})/P(N_{\rm m})$. Then,
\begin{align}
P(N_{\rm f}|N_{\rm m})&=C\sum^{\infty}_{N_{\rm t}=0}P(N_{\rm f}|N_{\rm t},N_{\rm m})P(N_{\rm m}|N_{\rm t}).
\end{align}
The last term in the above equation can be written as
\begin{align}
P(N_{\rm m}|N_{\rm t})&=\sum^{\infty}_{N_{\rm f'}=0}P(N_{\rm m},N_{\rm f'}|N_{\rm t})\\
                      &=\sum_{N_{\rm f'}=0}^{\infty}P(N_{\rm m}|N_{\rm f'},N_{\rm t})P(N_{\rm f'}|N_{\rm t})\\
                      &=\sum_{N_{\rm f'}=0}^{\infty}\delta(N_{\rm m}-(N_{\rm t}+N_{\rm f'}))P(N_{\rm f'}|N_{\rm t})\\
                      &=P(N_{\rm f'}=N_{\rm m}-N_{\rm t}|N_{\rm t}).
\end{align}
Using Eq.~\eqref{eq:binomial}, it follows
\begin{align}
 \begin{split}
 P(N_{\rm f}|N_{\rm m})={} & C\sum_{N_{\rm t}=0}^{\infty}P(N_{\rm f}|N_{\rm t},N_{\rm m})\binom{N_{\rm w}-N_{\rm t}}{N_{\rm m}-N_{\rm t}}\\
                           & [p(N_{\rm t})]^{N_{\rm m}-N_{\rm t}}[1-p(N_{\rm t})]^{N_{\rm w}-N_{\rm m}}              
\end{split}\\
\begin{split}
                       ={} & C\sum_{N_{\rm t}=0}^{\infty}\delta(N_{\rm m}-(N_{\rm t}+N_{\rm f}))\binom{N_{\rm w}-N_{\rm t}}{N_{\rm m}-N_{\rm t}}\\
                           & [p(N_{\rm t})]^{N_{\rm m}-N_{\rm t}}[1-p(N_{\rm t})]^{N_{\rm w}-N_{\rm m}}
\end{split}\\
\begin{split}\label{final_distribution}   
                       ={} & C\binom{N_{\rm w}-(N_{\rm m}-N_{\rm f})}{N_{\rm f}}[p(N_{\rm m}-N_{\rm f})]^{N_{\rm f}}\\
                           & [1-p(N_{\rm m}-N_{\rm f})]^{N_{\rm w}-N_{\rm m}}.
 \end{split}
\end{align}
Thus, the probability distribution of false matches obtained in Eq.~\eqref{final_distribution} appears identical to Eq.~\eqref{eq:binomial}, but with $N_{\rm t}$ replaced by $N_{\rm m}-N_{\rm f}$ (Eq.~\ref{eq:false_match1}) and a normalization constant $C$ that can be directly calculated.

It is then straightforward to calculate the expected number of false matches as
\begin{equation}
\left<N_{\rm f}\right>=\sum^{N_{\rm f}=N_{\rm m}}_{N_{\rm f}=0}N_{\rm f}P(N_{\rm f}|N_{\rm m}).
\label{eq:false_match2}
\end{equation}

\end{document}